\documentclass[a4paper,11pt]{article}
\usepackage{graphicx,epsfig}
\usepackage{fancyhdr,fancybox,float}
\usepackage{appendix}
\usepackage{titlesec}
\usepackage{etoolbox, chngcntr}
\usepackage{indentfirst}
\usepackage{verbatim}
\usepackage[sort&compress, numbers]{natbib}
\usepackage{geometry}
\usepackage{extarrows,chemarrow,xypic}
\usepackage{color}
\usepackage[small]{caption2}
\usepackage{microtype}
\DisableLigatures[f]{encoding = *, family = *}
\AtBeginEnvironment{appendices}{%
	\titleformat{\section}{\bfseries\large}{\appendixname~\thesection:}{0.5em}{}%
	\titleformat{\subsection}{\bfseries\large}{\thesubsection}{0.5em}{}%
}

\usepackage{times}                   

\usepackage{amssymb}                 
\usepackage{amsthm}
\usepackage{mathrsfs}                
\usepackage{bbm}
\usepackage{hyperref}                

\newcommand{\paperfont}{\fontsize{10pt}{1.1\baselineskip}\selectfont}
\pagebreak[4]

\begin{document}
\allowdisplaybreaks	
\theoremstyle{definition}
\makeatletter
\thm@headfont{\bf}
\makeatother

\newtheorem{theorem}{Theorem}
\newtheorem{definition}{Definition}
\newtheorem{lemma}{Lemma}
\newtheorem{proposition}{Proposition}
\newtheorem{corollary}{Corollary}
\newtheorem{remark}{Remark}
\newtheorem{example}{Example}
\newtheorem{assumption}{Assumption}


\lhead{}
\rhead{}
\lfoot{}
\rfoot{}

\renewcommand{\refname}{References}
\renewcommand{\figurename}{Figure}
\renewcommand{\tablename}{Table}
\renewcommand{\proofname}{Proof}
	
\newcommand{\diag}{\mathrm{diag}}
\newcommand{\tr}{\mathrm{tr}}
\newcommand{\re}{\mathrm{Re}}
\newcommand{\one}{\mathbbm{1}}
\newcommand{\Pnum}{\mathbb{P}}
\newcommand{\Enum}{\mathbb{E}}
\newcommand{\Rnum}{\mathbb{R}}
\newcommand{\dnum}{\mathrm{d}}
\newcommand{\hyper}{{}_2F_1}
\newcommand{\confl}{{}_1F_1}

\title{\textbf{Large deviations and fluctuation theorems for cycle currents defined in the loop-erased and spanning tree manners:\\ a comparative study}}
\author{Yuhao Jiang$^{1,\dag}$,\;\;\;Bingjie Wu$^{2,\dag}$,\;\;\;Chen Jia$^{1,*}$\\
\footnotesize $^1$ Applied and Computational Mathematics Division, Beijing Computational Science Research Center, Beijing 100193, China. \\
\footnotesize $^2$ LMAM, School of Mathematical Sciences, Peking University, Beijing 100871, China. \\
\footnotesize $^\dag$ These authors contribute equally to this work. \\
\footnotesize $^*$ Correspondence: chenjia@csrc.ac.cn}

\date{}
\maketitle
\thispagestyle{empty}

\paperfont

{\abstract
	The cycle current is a crucial quantity in stochastic thermodynamics. The absolute and net cycle currents of a Markovian system can be defined in the loop-erased (LE) or spanning tree (ST) manner. Here we make a comparative study between the large deviations and fluctuation theorems for LE and ST currents, i.e. cycle currents defined in the LE and ST manners. First, we derive the exact joint distribution and large deviation rate function for the LE currents of a system with a cyclic topology and also obtain the exact rate function for the ST currents of a general system. The relationship between the rate functions for LE and ST currents is clarified and the analytical results are applied to examine the fluctuations in the product rate of a three-step reversible enzyme reaction. Furthermore, we examine various types of fluctuation theorems satisfied by LE and ST currents and clarify their ranges of applicability. We show that both the absolute and net LE currents satisfy the strong form of all types of fluctuation theorems. In contrast, the absolute ST currents do not satisfy fluctuation theorems, while the net ST currents only satisfy the weak form of fluctuation theorems under the periodic boundary condition.}

\section{Introduction}
Over the past two decades, significant progress has been made in stochastic thermodynamics \cite{jarzynski2011equalities, seifert2012stochastic, van2015ensemble}, which has grown to become an influential branch of nonequilibrium statistical physics. In this field, a thermodynamic system is usually modelled by a Markov process. Markov chains, whose state spaces are discrete, are the most fundamental and important dynamic model since any Markov process can always be approximated by a Markov chain. Along this line, an equilibrium state is defined as a reversible Markov process and the deviation from equilibrium is usually quantified by the concept of entropy production, which can be represented as a bilinear function of thermodynamic fluxes and forces \cite{onsager1953fluctuations, hong2016novel}. It has long been noticed by Kolmogorov \cite{kolmogoroff1936theorie, jia2021detailed} that the reversibility of a Markov chain can be characterized by its cycle dynamics: the system is reversible if and only if the product of transition probabilities along each cycle and that along its reversed cycle are exactly the same, which generalizes the Wegscheider condition for detailed balanced chemical reaction networks. An incisive observation is that the entropy production can be decomposed along cycles, with the thermodynamics fluxes being the cycle currents (also called cycle fluxes or circulations) and with the thermodynamic forces being the cycle affinities \cite{schnakenberg1976network}.

The cycle representation theory of Markov chains has found wide applications in physics, chemistry, and biology \cite{zhang2012stochastic, ge2012stochastic}. In fact, the current of a cycle can be defined in several different ways. Two common definitions are based on the spanning tree and loop-erased methods. Hill \cite{hill1966studies3, hill1966studies4, hill1975free, hill1989free} and Schnakenberg \cite{schnakenberg1976network} developed a network theory and defined the currents for a family of \emph{fundamental cycles}. In this theory, a spanning tree is associated with the transition diagram of a Markov chain, which is a directed graph. Each edge of the graph that does not belong to the spanning tree, which is called a chord, will generate a fundamental cycle. The current of a fundamental cycle is defined as the number of times that the associated chord is traversed per unit time. The Qians \cite{qian1982circulation, qian1984circulations, jiang2004mathematical} further developed the cycle representation theory and defined the currents for all \emph{simple cycles} of the graph, i.e. cycles with no repeated vertices except the beginning and ending vertices. In this theory, the trajectory of a Markovian system is tracked. Once a cycle is formed, it is erased from the trajectory and we further keep track of the remaining trajectory until the next cycle is formed. The current of a simple cycle then is defined as the number of times that the cycle is formed per unit time. Recently, another type of cycle currents is proposed based on the idea of sequence matching \cite{pierpaolo2019exact, john2020reversal, patrick2021cycle}. In this theory, the currents are defined for \emph{all cycles} of the graph, i.e. directed paths with the first and last vertices being equal.

All types of cycle currents can also be defined along a single stochastic trajectory. One of the major advances in stochastic thermodynamics is the finding that a broad class of thermodynamic quantities such as entropy production and cycle currents satisfy various types of fluctuation theorems \cite{evans1993probability, gallavotti1995dynamical, jarzynski1997nonequilibrium, kurchan1998fluctuation, crooks1999entropy, searles1999fluctuation, lebowitz1999gallavotti, hatano2001steady, seifert2005entropy, esposito2010letter, lee2013fluctuation, ge2021martingale}, which provide nontrivial generalizations of the second law of thermodynamics in terms of equalities rather than inequalities. For cycle currents defined in the spanning tree manner, Andrieux and Gaspard \cite{andrieux2007fluctuation} proved that the fluctuation theorem holds for \emph{net} cycle currents in the long-time limit. Moreover, Polettini and Esposito \cite{polettini2014transient} showed that the transient fluctuation theorem at any finite time holds if the definition of cycle currents is slightly modified. For the cycle currents defined in the loop-erased manner, Andrieux and Gaspard \cite{andrieux2007network} and Jia et al. \cite{jia2016cycle} proved that all types of fluctuation theorems and symmetric relations are satisfied for both the \emph{absolute} and \emph{net} cycle currents. For cycle currents defined in the sequence matching manner, the corresponding fluctuation theorems and symmetric relations have also been developed recently \cite{patrick2021cycle}. The fluctuation theorems for cycle currents have also been developed for some stochastic processes with continuous state space, such as Langevin dynamics on circles \cite{ge2017cycle}.

From the mathematical perspective, another important question is whether various thermodynamic quantities defined along single stochastic trajectories satisfy the large deviation principle \cite{varadhan1984large, den2008large}. The large deviations are concerned with the long-time fluctuation behavior of a stochastic process with small probability and it is closed related to the fluctuation theorem in the long-time limit. For Markovian systems, the large deviations for empirical measures, i.e. the number of times that each vertex of the graph is crossed per unit time, and for empirical flows, i.e. the number of times that each edge of the graph is traversed per unit time, have been extensively studied, while the large deviations for empirical cycle currents, i.e. the number of times that each cycle of the graph is formed per unit time, have received comparatively little attention. For cycle currents defined in the spanning tree manner, the large deviations have been established since in this case, the empirical cycle currents are exactly the empirical flows of chords \cite{bertini2015flows, bertini2015large}. For cycle currents defined in the loop-erased manner, the explicit expression of the large deviation rate function is still unknown, even for systems with a simple topological structure.

In this paper, we make a comprehensive comparative study between cycle currents defined in the spanning tree and loop-erased manners, and clarify the connections and differences between them. The structure of this paper is organized as follows. In Section 2, we recall the definitions of the two types of cycle currents and make a brief comparison between them. In Section 3, we investigate the large deviations for the two types of cycle currents. We obtain the exact joint distribution and rate function for loop-erased currents of a monocyclic Markovian system using the so-called cycle insertion method, and also obtain the exact rate function for spanning tree currents of a general Markovian system. In Section 4, we state and compare various types of fluctuation theorems and symmetric relations satisfied by the two types of cycle currents. We clarify the ranges of applications of these fluctuation theorems and show that all the results for spanning tree currents can be derived naturally from the relevant results for loop-erased currents. We conclude in Section 5.

\section{Model and two types of cycle currents}

\subsection{Model}
\begin{figure}[htb!]
	\centering\includegraphics[width=0.7\textwidth]{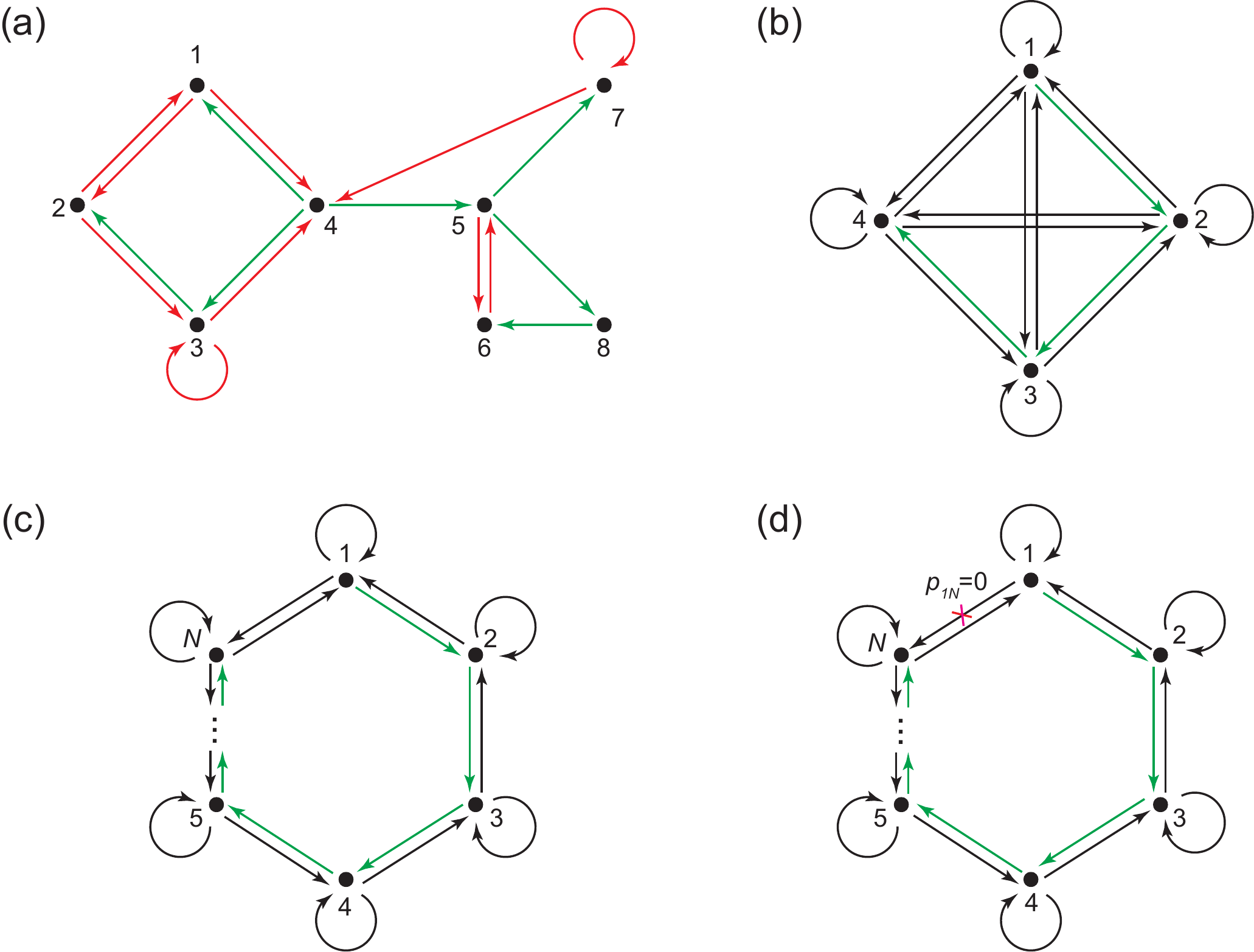}
	\caption{\textbf{Transition diagrams and the associated spanning trees for various Markov chains.} (\textbf{a}) A Markov chain with a general transition diagram. The green arrows show the spanning tree $T$ with root vertex $4$, and the red arrows show all the chords of $T$. (\textbf{b}) A fully connected Markov chain with four states, where each state can transition to both itself and any other states. (\textbf{c}) A monocyclic Markov chain with $N$ states. Each state can only transition to itself and its two neighbours. (\textbf{d}) A monocyclic Markov chain with $N$ states. Here the transition from state $1$ to state $N$ is forbidden. In (b)-(d), the green arrows show the spanning tree $T$.}\label{figure:transitiongraph}
\end{figure}

Here we consider a thermodynamic system modelled by a discrete-time Markov chain $\xi=(\xi_n)_{n\ge 0}$ with state space $S=\{1,2,\cdots,N\}$ and transition probability matrix $P = (p_{ij})_{i,j\in S}$, where $p_{ij}$ denotes the transition probability from state $i$ to state $j$. The transition diagram of the Markov chain $\xi$ is a directed graph $G = (S,E)$, where the vertex set $S$ is the state space and the edge set $E$ contains all directed edges with positive transition probabilities (Fig. \ref{figure:transitiongraph}). In this paper, we use $\langle i, j\rangle$ to denote the edge from state $i$ to state $j$. With this notation, the edge set $E$ can be written more clearly as
\begin{equation*}
	E = \{\langle i,j\rangle\in S\times S: p_{ij}>0\},
\end{equation*}
and we assume that $|E| = M$, where $|E|$ denotes the number of elements in $E$. Here we assume that the Markov chain $\xi$ is irreducible, which means that $G$ is a connected graph. Since the transition from a particular state to itself is allowed for a Markov chain, the graph $G$ may contain an edge from a state to itself, i.e. a self-loop (Fig. \ref{figure:transitiongraph}).

A special case occurs when the transition diagram $G$ has a cyclic topology (except all self-loops), as illustrated in Fig. \ref{figure:transitiongraph}(c). Such systems will be referred to as monocyclic Markov chains in this paper. Specifically, the Markov chain $\xi$ is called \emph{monocyclic} if $p_{ij}=0$ for any $|i-j|\ge 2$, where $i$ and $j$ are understood to be modulo $N$. In fact, monocyclic systems are of particular relevance in the biological context. Many crucial biochemical processes such as conformational changes of enzymes and ion channels \cite{cornish2012fundamentals, sakmann1995single}, progression of cell cycle \cite{jia2021frequency, jia2021cell}, phenotypic switching of cell types \cite{gupta2011stochastic, jia2014modeling}, phosphorylation-dephosphorylation cycle \cite{qian2007phosphorylation, beard2008chemical}, and activation of promoters due to chromatin remodeling and transcription factor binding \cite{pedraza2008effects, jia2022analytical} can all be modelled as monocyclic Markov chains. In what follows, we mainly focus on monocyclic systems, while most of the results can be extended to general systems.

\subsection{Cycle currents defined in the loop-erased manner}
In this paper, we will investigate and compare two types of cycle currents. We first recall cycle currents defined in the loop-erased manner \cite{jiang2004mathematical, kalpazidou2007cycle}. A \emph{circuit} of the Markov chain $\xi$ is defined as a path $i_1 \to i_2 \to\cdots\to i_s \to i_1$ in the graph $G$ from a state to itself, where $i_1,i_2,\cdots,i_s$ are distinct states in $S$. Let $j_1 \to j_2 \to\cdots\to j_r \to j_1$ be another circuit. The above two circuits are said to be \emph{equivalent} if $r = s$ and there exists an integer $k$ such that
\begin{equation*}
	j_1 = i_{k+1},j_2 = i_{k+2},\cdots,j_r = i_{k+s},
\end{equation*}
where $k+1,\cdots,k+s$ are understood to be modulo $s$. The equivalence class of the circuit $i_1 \to i_2 \to\cdots\to i_s \to i_1$ under the equivalence relation described above is called a \emph{cycle} and is often denoted by $c = (i_1,i_2,\cdots,i_s)$. For example, $(1,2,3)$, $(2,3,1)$ and $(3,1,2)$ represent the same cycle. The \emph{reversed cycle} of $c = (i_1,i_2,\cdots,i_s)$ is defined as $c- = (i_1,i_s,\cdots,i_2)$. The set of all cycles is called the \emph{cycle space} and is denoted by $\mathcal{C}$.

The trajectory of a Markov chain constantly forms various cycles. Intuitively, if we discard the cycles formed by $\xi$ and keep track of the remaining states in the trajectory, then we obtain a new Markov chain $\tilde{\xi} = (\tilde{\xi}_n)_{n\geq 0}$ called the \emph{derived chain}. For example, if the trajectory of the original chain $\xi$ is $\{1,2,3,3,2,3,4,1,4,\cdots\}$, then the corresponding trajectory of the derived chain $\tilde{\xi}$ and the cycles formed are shown in Table \ref{trajectory}.
\begin{table}[htb!]
	{\small\renewcommand\arraystretch{1.3}\centering
		\begin{tabular}{cccccccccc} \hline\hline
			$n$             & 0 & 1 & 2 & 3   & 4     & 5 & 6 & 7         & 8 \\ \hline
			$\xi_n$          & 1 & 2 & 3 & 3   & 2     & 3 & 4 & 1         & 4 \\ \hline
			$\tilde{\xi}_n$ & {[}1{]} & {[}1,2{]} & {[}1,2,3{]} & {[}1,2,3{]} & {[}1,2{]} & {[}1,2,3{]} & {[}1,2,3,4{]} & {[}1{]} & {[}1,4{]} \\ \hline
			cycles formed &   &   &   & (3) & (2,3) &   &   & (1,2,3,4) &   \\ \hline\hline
		\end{tabular}
		\caption{\textbf{An example of the derived chain and the cycles formed.}}\label{trajectory}}
\end{table}

Moreover rigorously, a state of the derived chain $\tilde{\xi}$ is a finite sequence $i_1,i_2,\cdots,i_s$ of distinct states in $S$, denoted by $[i_1,i_2,\cdots,i_s]$. Suppose that $\tilde{\xi}_{n-1}=[i_1,i_2,\cdots,i_s]$ and $\xi_n = i_{s+1}$. If $i_{s+1}\notin\{i_1,i_2,\cdots,i_s\}$, then $\tilde{\xi}_n$ is defined as (see Table \ref{figure:transitiongraph} for an illustration)
\begin{equation*}
	\tilde{\xi}_n = [i_1,i_2,\cdots,i_s,i_{s+1}].
\end{equation*}
On the other hand, if $i_{s+1}=i_r$ for some $1\le r\le s$, then $\tilde{\xi}_n$ is defined as (see Table  \ref{figure:transitiongraph} for an illustration) \begin{equation*}
	\tilde{\xi}_n = [i_1,i_2,\cdots,i_r].
\end{equation*}
In this case, we say that the Markov chain $\xi$ forms cycle $c = (i_r,i_{r+1},\cdots,i_s)$ at time $n$. Let $N^c_n$ be the number of times that cycle $c$ is formed up to time $n$. Then the \emph{empirical (absolute) current} of cycle $c$ up to time $n$ is defined as
\begin{equation*}
	J_n^c = \frac{1}{n}N^c_n,
\end{equation*}
and the \emph{empirical net current} of cycle $c$ up to time $n$ is defined as $\tilde{J}^c_n = J^c_n-J^{c-}_n$. Intuitively, $J^c_n$ represents the number of times that cycle $c$ is formed per unit time and $\tilde{J}^c_n$ represents the net number of times that cycle $c$ is formed per unit time.

As $n\rightarrow\infty$, the empirical cycle current $J^c_n\rightarrow J^c$ and empirical net cycle current $\tilde{J}^c_n\rightarrow\tilde{J}^c$ will both converge with probability one. The limits $J^c$ and $\tilde{J}^c$ are called the \emph{current} and \emph{net current} of cycle $c$, respectively. The explicit expressions of $J_c$ and $\tilde{J}_c$ can be found in \cite{jiang2004mathematical}. The well-known cycle current decomposition theorem \cite{jiang2004mathematical} states that
\begin{equation}\label{decomposition}
	\pi_ip_{ij} = \sum_{c\ni\langle i,j\rangle}J^c,
\end{equation}
where $\pi_i$ is the steady-state probability of state $i$ and the sum on the right-hand side is taken over all cycles $c$ which traverses edge $\langle i,j\rangle$ (the symbol $c\ni\langle i,j\rangle$ means that cycle $c$ traverses edge $\langle i,j\rangle$). This shows that the probability flux between any pair of states can be decomposed as the sum of cycle currents.

\subsection{Cycle currents defined in the spanning tree manner}\label{subsection:spanning tree}
The current of a cycle can also be defined in the spanning tree manner \cite{schnakenberg1976network, kalpazidou2007cycle}. Let $T$ be a directed subgraph of the transition diagram $G$, i.e. all the edges of $T$ are also edges of $G$, and let $\overline{T}$ denote the undirected graph associated with $T$. Recall that $T$ is called a \emph{spanning tree} (or maximal tree) of $G$ if the following three conditions are satisfied \cite{kalpazidou2007cycle}:
\begin{itemize}
	\item[(a)] $T$ is a covering subgraph of $G$, i.e. $T$ contains all the vertices of
	$G$;
	\item[(b)] $\overline{T}$ is connected;
	\item[(c)] $\overline{T}$ has no circuits, where a circuit of an undirected graph is defined as an undirected path from a vertex to itself.
\end{itemize}
In the following, we use $T$ to represent both the spanning tree itself and its edge set. The meaning should be clear from the context. In general, the choice of the spanning tree is not unique, which means that a graph may have many different spanning trees. It is easy to see that any spanning tree $T$ must contain all the vertices of $G$ and must have $N-1$ edges (see the green arrows in Fig. \ref{figure:transitiongraph}) \cite{kalpazidou2007cycle}.

A directed edge $l\notin T$ is called a \emph{chord} of $T$ (see the red arrows in Fig. \ref{figure:transitiongraph}(a)). Since $|E| = M$ and $|T| = N-1$, any spanning tree $T$ must have $M-N+1$ chords. Since $\overline{T}$ is connected and has no circuits, if we add to $T$ one of its chord $l$, then the resulting undirected subgraph $\overline{T \cup \{l\}}$ must have exactly one circuit. Let $c_l$ be the cycle obtained from this circuit with the orientation being the same as the chord $l$. For example, for the system illustrated in Fig. \ref{figure:transitiongraph}(a), if we add the chord $l = \langle 2,1\rangle$ to the spanning tree $T$, then we obtain the cycle $c_l = (2,1,4,3)$. The family of cycles $\mathcal{L} = \{c_l: l\notin T\}$ generated by the chords is referred to as the \emph{fundamental set}. Since there is a one to one correspondence between the chord set and the fundamental set, the number of times that cycle $c_l$ is formed is simply defined as the number of times that chord $l$ is traversed. Along this line, the \emph{empirical (absolute) current} of cycle $c_l$ up to time $n$ is defined as
\begin{equation*}
	Q^{c_l}_n = \frac{1}{n}\sum_{m=1}^n1_{\{\langle\xi_{m-1},\xi_m\rangle=l\}}.
\end{equation*}
Intuitively, $Q^{c_l}_n$ represents the number of times that chord $l$ is traversed per unit time. Unlike the loop-erased technique which can be used to define the currents of all cycles, the spanning tree technique can only be used to define the currents of cycles in the fundamental set.

Similarly, we can define the empirical net current in the spanning tree manner. The \emph{empirical net current} of cycle $c_l$ up to time $n$ is defined as $\tilde{Q}^{c_l}_n=Q^{c_l}_n-Q^{c_l-}_n$. If $c_l$ is composed of one or two states, then $c_l = c_l-$ and thus $\tilde{Q}^{c_l}_n = 0$. For any chord $l=\langle i,j\rangle$, if $c_l$ is composed of three or more states and if $c_l-$ is in the fundamental set, then $l-=\langle j,i\rangle$ must also be a chord and $c_l-$ is exactly the cycle generated by the chord $l-$. As $n\rightarrow\infty$, the empirical cycle current $Q^{c_l}_n\rightarrow Q^{c_l}$ and empirical net cycle current $\tilde{Q}^{c_l}_n\rightarrow\tilde{Q}^{c_l}$ will both converge with probability one. The limits $Q^{c_l}$ and $\tilde{Q}^{c_l}$ are called the \emph{current} and \emph{net current} of cycle $c_l$, respectively. For any chord $l=\langle i,j\rangle$, it follows from the ergodic theorem of Markov chains that $Q^{c_l} = \pi_ip_{ij}$.

We emphasize that most previous papers focused on net cycle currents defined in the loop-erased \cite{andrieux2007network} and spanning tree \cite{schnakenberg1976network, andrieux2007fluctuation} manners, and absolute cycle currents have received much less attention. Clearly, the net currents vanish for any one-state and two-state cycles. Hence in previous papers \cite{schnakenberg1976network, andrieux2007fluctuation, andrieux2007network}, the net currents are only defined for cycles with three or more states. In this paper, we focus on both absolute and net currents. Here, following \cite{kalpazidou2007cycle,jiang2004mathematical}, we extend the definition slightly to include cycles with one or two states. This extension turns out to be useful, as can be seen in Section \ref{sec:joint} below.

\subsection{Comparisons between two types of cycle currents}\label{comparison}
Next we make a brief comparison between the two types of cycle currents. In what follows, cycle currents defined in the loop-erased manner will be called \emph{LE currents} and those defined in the spanning tree manner will be called \emph{ST currents}. We have seen that LE currents are defined for all cycles in the cycle space $\mathcal{C}$, while ST currents are only defined for cycles in the fundamental set $\mathcal{L}$. Hence LE currents provide a more complete description of the cycle dynamics than ST currents. Moreover, since the spanning tree is in general not unique, different choices of the spanning tree correspond to different ST currents. Clearly, LE currents are independent of the choice of the spanning tree.

A natural question is how much the fundamental set $\mathcal{L}$ is smaller than the cycle space $\mathcal{C}$. Since each chord corresponds one and only one element in $\mathcal{L}$, we have $|\mathcal{L}| = M-N+1$. It is difficult to provide a unified expression for $|\mathcal{C}|$. To gain deeper insights, we focus on two special cases. We first consider a Markov chain whose transition diagram is fully connected, i.e. $p_{ij}>0$ for any $i,j\in S$, as illustrated in Fig. \ref{figure:transitiongraph}(b). In this case, the number of cycles with $k$ states is given by $N(N-1)\cdots (N-k+1)/k$, and thus
\begin{equation*}
	|\mathcal{C}| = \sum_{k=1}^N\frac{N(N-1)\cdots (N-k+1)}{k}.
\end{equation*}
In particular, when $N = 4$, we have $|\mathcal{C}| = 24$ and the cycle space is given by
\begin{align*}
	\mathcal{C} = \{&(1),(2),(3),(4),(1,2),(1,3),(1,4),(2,3),(2,4),(3,4),\\
	&(1,2,3),(1,2,4),(1,3,2),(1,3,4),(1,4,2),(1,4,3),(2,3,4),(2,4,3),\\
	&(1,2,3,4),(1,2,4,3),(1,3,2,4),(1,3,4,2),(1,4,2,3),(1,4,3,2)\}.
\end{align*}
If we choose the spanning tree to be $T = 1\to 2\to 3\to 4$, then $|\mathcal{L}|=13$ and the fundamental set is given by
\begin{align*}
	\mathcal{L} = \{&(1),(2),(3),(4),(1,2),(2,3),(3,4)\\
	&(1,2,3),(1,3,2),(2,3,4),(2,4,3),(1,2,3,4),(1,4,3,2)\}.
\end{align*}
For a fully connected system, the number of ST currents is much smaller than the number of LE currents.

We next consider the monocyclic Markov chain illustrated in Fig. \ref{figure:transitiongraph}(c), where each state can only transition to itself and its two neighbours. In this case, we have $|\mathcal{C}|=2N+2$ and the cycle space is given by
\begin{equation}\label{def:monocyclic C}
	\mathcal{C} = \{(1),\cdots,(N),(1,2),\cdots,(N-1,N),(N,1),(1,2,\cdots,N),(1,N,\cdots,2)\}.
\end{equation}
The first $N$ cycles are one-state cycles, i.e. self-loops, the middle $N$ cycles are two-state cycles, and the last two cycles are $N$-state cycles. If we choose the spanning tree to be $T = 1\to 2\to\cdots \to N$, then $|\mathcal{L}|=2N+1$ and the fundamental set is given by
\begin{equation*}
	\mathcal{L} = \{(1),\cdots,(N),(1,2),\cdots,(N-1,N),(1,2,\cdots,N),(1,N,\cdots,2)\}.
\end{equation*}
For a monocyclic system, there is only one cycle, i.e. cycle $(N,1)$, that is contained in $\mathcal{C}$ but is not contained in $\mathcal{L}$.

To further understand the relationship between the LE current $J^c_n$ and the ST current $Q^{c_n}_n$, we use the convention of periodic boundary condition, i.e. $\xi_0=\xi_n$, which is a standard assumption in the literature \cite{den2008large}. With this assumption, for any chord $l$, it is easy to see that
\begin{equation}\label{conversion}
	Q_n^{c_l} = \sum_{c\ni l}J^c_n,
\end{equation}
where the sum is taken over all cycles $c$ that traverse chord $l$. Both sides of the equation represent the number of times that chord $l$ is formed per unit time. This shows that ST currents can be represented as the sum of LE currents.

\section{Joint distribution and large deviations for cycle currents}\label{sec:joint}
Previous studies about cycle currents mainly focused on the fluctuation relations, i.e. the symmetry relations satisfied by the probability distribution of cycle currents \cite{jia2016cycle,andrieux2007network}. However, very little is known about the explicit expression of the probability distribution. Here we will address this problem and then use it to study the large deviations for cycle currents. In Section \ref{LELDP}, we use methods in combinatorics and graph theory to compute the explicit expression of the joint probability distribution for LE currents. In Section \ref{sec:LDP}, using the exact joint distribution and the Stirling formula, we investigate the large deviations for LE currents and give the explicit expression of the corresponding rate function. In Section \ref{FDTST}, we study the large deviations for ST currents using the existing large deviation results for empirical flows.

\subsection{Joint distribution for LE currents of monocyclic Markov chains}\label{LELDP}
We first focus on the joint distribution for empirical LE currents $(J^c_n)_{c\in\mathcal{C}}$. In general, it is very difficult to obtain the explicit expression of the joint distribution for a general Markov chain. Here we focus on the monocyclic system illustrated in Fig. \ref{figure:transitiongraph}(c). All possible cycles formed by the system are listed in \eqref{def:monocyclic C}. Without loss of generality, we assume that the system starts from state $1$. For each cycle $c = (i_1,i_2,\cdots,i_s)$, let $\gamma^c = p_{i_1i_2}p_{i_2i_3}\cdots p_{i_si_1}$ denote the product of transition probabilities along this cycle. For any sequence of negative integers $k=(k^c)_{c\in\mathcal{C}}$ satisfying $\sum_{c\in\mathcal{C}}|c|k^c = n$, since we have assumed the periodic boundary condition, the joint distribution of empirical LE currents is given by
\begin{equation*}
	\mathbb{P}\left(J^c_n=\nu^c,\;\forall c\in\mathcal{C}\right)
	=\mathbb{P}\left(N^c_n=k^c,\;\forall c\in\mathcal{C}\right) =|G_n(k)|\prod_{c\in\mathcal{C}}\left(\gamma^c\right)^{k^c},
\end{equation*}
where $\nu^c = k^c/n$ is the frequency of occurrence of cycle $c$ and $G_n(k)$ denotes the set of all possible trajectories up to time $n$ so that each cycle $c$ is formed $k^c$ times. Such trajectories will be called \emph{allowable trajectories} in what follows. For convenience, we write $k^c$ as $k^i$ if $c=(i)$ is a one-state cycle, as $k^{i,i+1}$ if $c=(i,i+1)$ is a two-state cycle, as $k^+$ if $c=(1,2,\cdots,N)$ is the clockwise $N$-state cycle, and as $k^-$ if $c=(1,N,\cdots,2)$ is the counterclockwise $N$-state cycle (Fig. \ref{figure:transitiongraph}(c)). For example, for a three-state system, if the sequence $k = (k^c)_{c\in\mathcal{C}}$ is chosen as
\begin{equation}\label{sequence}
	k^3 = k^{12} = k^{23} = k^- = 1,\;\;\;k^1 = k^2 = k^{13} = k^+= 0,
\end{equation}
then there are eight allowable trajectories up to time $n = 8$, and all of them are listed in Table \ref{table:all possible trajectories}. Similarly, we write $\nu^c$ as $\nu^i$, $\nu^{i,i+1}$, $\nu^+$, and $\nu^-$, and write $J^c$ as $J^i$, $J^{i,i+1}$, $J^+$, and $J^-$.
\begin{table}[htb!]
	\renewcommand\arraystretch{1.2}
	\begin{tabular}{cccccccccc}
		\hline\hline
		$m$    & 0 & 1 & 2 & 3 & 4 & 5 & 6 & 7 & 8 \\ \hline
		$\xi_m$ & 1 & 3 & 3 & 2 & 3 & 2 & 1 & 2 & 1 \\ \hline
		$\xi_m$ & 1 & 3 & 2 & 3 & 3 & 2 & 1 & 2 & 1 \\ \hline
		$\xi_m$ & 1 & 3 & 3 & 2 & 1 & 2 & 3 & 2 & 1 \\ \hline
		$\xi_m$ & 1 & 3 & 2 & 1 & 2 & 3 & 3 & 2 & 1 \\ \hline
		$\xi_m$ & 1 & 2 & 3 & 3 & 2 & 1 & 3 & 2 & 1 \\ \hline
		$\xi_m$ & 1 & 2 & 3 & 2 & 1 & 3 & 3 & 2 & 1 \\ \hline
		$\xi_m$ & 1 & 2 & 1 & 3 & 3 & 2 & 3 & 2 & 1 \\ \hline
		$\xi_m$ & 1 & 2 & 1 & 3 & 2 & 3 & 3 & 2 & 1 \\ \hline\hline
	\end{tabular}\centering
	\caption{\textbf{An example of allowable trajectories for a monocyclic system.} All the eight allowable trajectories for a three-state system up to time $n = 8$ so that each one of the four cycles $(3)$, $(12)$, $(23)$, and $(1,3,2)$ is formed once, while the remaining four cycles $(1)$, $(2)$, $(13)$, and $(1,2,3)$ are not formed, i.e. $k^3 = k^{12} = k^{23} = k^- = 1$ and $k^1 = k^2 = k^{13} = k^+= 0$.}\label{table:all possible trajectories}
\end{table}

We will next compute the number $|G_n(k)|$ of allowable trajectories. The basic idea is to insert all cycles into the trajectory in some appropriate order. The number of all possible insertions will then be the number of all allowable trajectories. The calculation is divided into the following three steps.

\begin{figure}[htb!]
	\centering\includegraphics[width=1.0\textwidth]{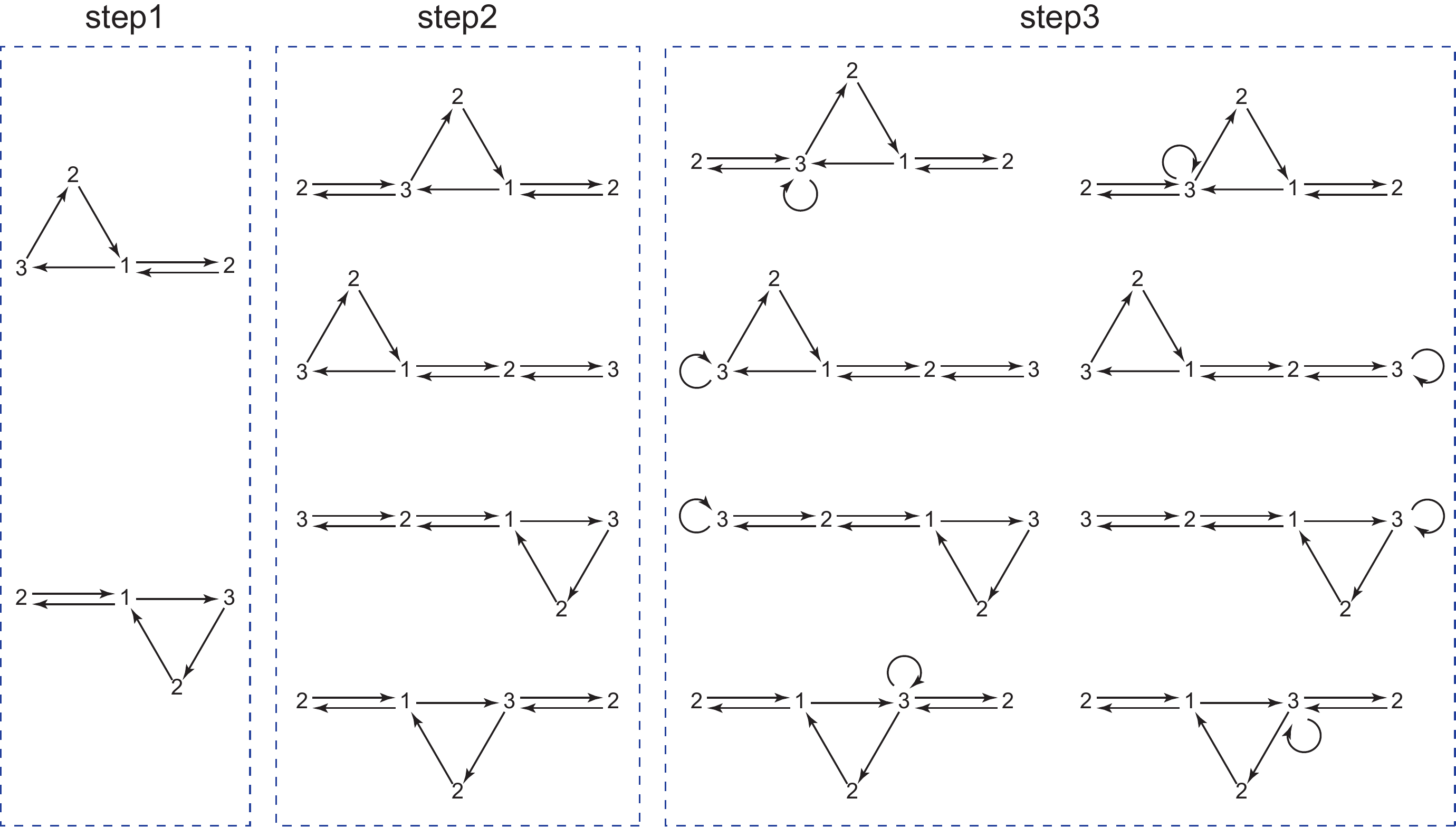}
	\caption{\textbf{Schematic of the cycle insertion method of constructing all allowable trajectories.} Here we use the example given in \eqref{sequence}. The cycle insertion method is divided into three steps: first we insert all the cycles containing the initial state into the trajectory, next we insert all the remaining two-state cycles into the trajectory, and finally we insert all the remaining one-state cycles into the trajectory. After the three-step cycle insertion, we find all the eight allowable trajectories, which coincide exactly with those listed in Table \ref{table:all possible trajectories}.}\label{figure:insertion}
\end{figure}

1) Since we have assumed that the system starts from state 1, as the first step, we select all the cycles containing the initial state 1, i.e. $(1)$, $(1,2)$, $(N,1)$, $(1,2,\cdots,N)$, $(1,N,\cdots,2)$, and insert them into the trajectory. Since each cycle $c$ is formed $k^c$ times, the total number of possible insertions in step 1), i.e. the number of all permutations of these cycles, are given by	
\begin{equation*}\label{formula:A1}
	A_1 = \binom{k^1+k^{12}+k^{N1}+k^{+}+k^{-}}{k^1,k^{12},k^{N1},k^{+},k^{-}}
	:= \frac{(k^1+k^{12}+k^{N1}+k^{+}+k^{-})!}{k^1!\;k^{12}!\;k^{N1}!\;k^{+}!\;k^{-}!}.
\end{equation*}
For the example given in \eqref{sequence}, all possible insertions in step 1) are shown in the left panel of Fig. \ref{figure:insertion}.

2) We next insert the remaining two-state cycles $(2,3),(3,4),\cdots,(N-1,N)$ into the trajectory. Note that when the system forms a two-state cycle $(i,i+1)$, it may be formed at state $i$ or state $i+1$. For example, for the trajectory $\{1,3,2,3,\cdots\}$, when cycle $(2,3)$ is formed, the derived chain becomes $[1,3]$. In this case, we say that the cycle is formed at state $3$. On the contrary, for the trajectory $\{1,2,3,2,\cdots\}$, when cycle $(2,3)$ is formed, the derived chain becomes $[1,2]$. In this case, we say that the cycle is formed at state $2$.

For any two-state cycle $(i,i+1)$, let $l^{i}$ and $m^{i}$ denote the number of times that it is formed at state $i$ and state $i+1$, respectively. Clearly, we have $l^{i}+m^{i} = k^{i,i+1}$. When $l^i$ and $m^i$ are fixed, the number of allowable trajectories can be computed as follows. First we insert the $l^2$ cycle $(2,3)$ at state $2$. There are $k^{12}+k^+$ possible positions for the insertion, which correspond to state $2$ in the cycles $(1,2)$ and $(1,2,\cdots, N)$, which have been arranged in step 1). Note that these positions do not include state $2$ in the cycle $(1,N,\cdots,2)$. This is because if we insert cycle $(2,3)$ here, then the cycle will be formed at state $3$ rather than state $2$. Hence the number of possible insertions is given by
\begin{equation}\label{p1}
	\binom{k^{12}+k^++l^{2}-1}{l^{2}}.
\end{equation}
Then we insert the $l^{i}$ cycle $(i,i+1)$ at state $i$ one by one for $3\leq i\leq N-1$. For each $i$, there are $l^{i-1}+k^+$ possible positions for the insertion, which correspond to state $i$ in the cycles $(i-1,i)$ and $(1,2,\cdots,N)$. The number of possible insertions is given by
\begin{equation}\label{p2}
	\binom{l^{i-1}+k^++l^{i}-1}{l^{i}},\;\;\;3\le i\le N-1.
\end{equation}
Thus far, we have inserted the $l^{i}$ cycle $(i,i+1)$ at state $i$ one by one for $2\leq i\leq N-1$. Combining \eqref{p1} and \eqref{p2}, the number of possible insertions is given by
\begin{equation*}
	\prod_{i=2}^{N-1}\binom{l^{i}+l^{i-1}+k^{+}-1}{l^{i}},
\end{equation*}
where $l^{1} := k^{12}$.

Next we insert the $m^{i}$ cycle $(i,i+1)$ at state $i+1$ one by one for $2\leq i\leq N-1$ in a similar way, and the number of possible insertions is given by
\begin{equation*}
	\prod_{i=2}^{N-1}\binom{m^{i}+m^{i+1}+k^{-}-1}{m^{i}},
\end{equation*}
where $m^{N} := k^{N1}$. Up till now, we have inserted all the two-state cycles into the trajectory. Summing over all choices of $l^{i}$ and $m^{i}$, the total number of possible insertions in step 2) is given by
\begin{equation*}
	A_2 = \sum_{l^{2}+m^{2}=k^{23}}\dots\sum_{l^{N-1}+m^{N-1}=k^{N-1,N}}
	\prod_{i=2}^{N-1}\binom{l^{i}+l^{i-1}+k^{+}-1}{l^{i}}\prod_{i=2}^{N-1}\binom{m^{i}+m^{i+1}+k^{-}-1}{m^{i}}.
\end{equation*}
For the example given in \eqref{sequence}, all possible insertions in step 2) are shown in the middle panel of Fig. \ref{figure:insertion}.

3) We finally insert the remaining one-state cycles into the trajectory. Specifically, we insert cycle $(i)$ into the trajectory one by one for $2\leq i\leq N$. For each $i$, there are $\sum_{c\ni i}k^c-k^i$ possible positions for the insertion, which correspond to state $i$ in all the cycles except cycle $(i)$. Hence the total number of possible insertions in step 3) is given by
\begin{equation*}\label{formula:A3}
	A_3 = \prod_{i=2}^N\binom{\sum_{c\ni i}k^{c}-1}{k^{i}}.
\end{equation*}
For the example given in \eqref{sequence}, all possible insertions in step 3) are shown in the right panel of Fig. \ref{figure:insertion}.

Combining the above three steps, we finally obtain the number of allowable trajectories, which is given by
\begin{equation*}
	|G_n(k)|=A_1A_2A_3.
\end{equation*}
Hence the joint distribution of empirical LE currents can be computed exactly as
\begin{equation}\label{joint}
	\mathbb{P}\left(J^c_n=\nu^c,\;\forall c\in\mathcal{C}\right)
	= A_1A_2A_3\prod_{c\in\mathcal{C}}\left(\gamma^c\right)^{k^c}.
\end{equation}
We have seen that most previous papers \cite{schnakenberg1976network, andrieux2007fluctuation, andrieux2007network} mainly focus on cycles with three or more states since the net currents for all one-state and two-state cycles must vanish. Here, we extend the definition slightly to include cycles with one and two states. This extension has the following two advantages: (i) in this paper, we not only focus on net cycle currents but also focus on absolute cycle currents; it is clear that the absolute currents for one-state and two-state cycles do not vanish and thus cannot be ignored; (ii) only when all one-state and two-state cycles are taken into account, it is possible to recover all the allowable trajectories from empirical cycle currents using the three-step cycle insertion method; in this way, the joint distribution of empirical cycle currents has a simple closed-form expression.

\subsection{Large deviations for LE currents of monocyclic Markov chains}\label{sec:LDP}
The large deviations are concerned with the long-time fluctuation behavior of a stochastic process with small probability \cite{varadhan1984large, den2008large}. We next investigate the large deviations for empirical LE currents of a monocyclic Markov chain. Note that under the periodic boundary condition, the empirical LE currents $(J^c_n)_{c\in \mathcal{C}}$ must lie in the space
	\begin{equation*}
		\mathcal{V} = \left\{(\nu^c)_{c\in \mathcal{C}}:\;\nu^c\geq 0,\;
		\sum_{c\in \mathcal{C}}|c|\nu^c  = 1\right\},
	\end{equation*}
	where $|c|$ denotes the length of cycle $c$, i.e. the number of states contained in cycle $c$. Roughly speaking, $(J^c_n)_{c\in\mathcal{C}}$ are said to satisfy a large deviation principle with rate function $I_J:\mathcal{V}\rightarrow[0,\infty]$ if the joint distribution satisfies
	\begin{equation}\label{LDP}
		\mathbb{P}(J^c_n=\nu^c,\;\forall c\in\mathcal{C})\propto e^{-n I_J(\nu)},\;\;\;n\to\infty,
	\end{equation}
	for any $\nu = (\nu^c)_{c\in\mathcal{C}}\in\mathcal{V}$. Clearly, the large deviation theory can capture the long-time fluctuation behavior of cycle currents. Next we only present the main idea of the proof. The rigorous definition and proof of the large deviation principle can be found in Section 1 of Supplementary Material.

To obtain the explicit expression of the rate function $I_J$, we recall the Stirling formula
\begin{equation*}
	\log n! = n\log n-n+O(\log n)=h(n)-n+O(\log n),
\end{equation*}
where $h(x) = x\log x$ for any $x\geq 0$. For convenience, set $k_i=\sum_{c\ni i}k^c$ and $\nu_i=\sum_{c\ni i}\nu^c$. Note that the definitions of $k_i$ and $k^i$ are different. It then follows from the Stirling formula that
\begin{equation}\label{log A1}
	\begin{split}
		\log A_1 &= \log\frac{k_1!}{k^1!\;k^{12}!\;k^{N1}!\;k^+!\;k^-!}\\
		&= h(k_1)-h(k^1)-h(k^{12})-h(k^{N1})-h(k^+)-h(k^-)+O(\log n)\\
		&= n\left[h(\nu_1)-h(\nu^1)-h(\nu^{12})-h(\nu^{N1})-h(\nu^+)-h(\nu^-)\right]+O(\log n).
	\end{split}
\end{equation}
Similarly, we have
\begin{equation}\label{log A3}
	\begin{split}
		\log A_3&=\log\prod_{i=2}^N\binom{k_i-1}{k^{i}}
		=\sum_{i=2}^N\log\frac{k_i!}{k^i!\left(k_i-k^i\right)!}\\
		&=\sum_{i=2}^N\left[h(k_i)-h(k^i)-h(k_i-k^i)\right]+O(\log n)\\
		&=\sum_{i=2}^Nn\left[h(\nu_i )-h(\nu^i)-h(\nu_i-\nu^i)\right]+O(\log n).
	\end{split}
\end{equation}
Finally, we estimate $\log A_2$. Let $D = \{(l^i,m^i)_{2\le i\le N-1}:\;l^i,m^i\in\mathbb{N},\;l^i+m^i=k^{i,i+1}\}$ denote the set of all possible choices of $l^i$ and $m^i$. For any $L = (l^i,m^i)\in D$, let
\begin{equation*}
	B_L=\prod_{i=2}^{N-1}\binom{l^{i}+l^{i-1}+k^{+}-1}{l^{i}}\binom{m^{i}+m^{i+1}+k^{-}-1}{m^{i}}
\end{equation*}
be the number of insertions in step 2) when $l^i$ and $m^i$ are fixed. It is clear that $|D|\le n^{N-2}$. Thus we have
\begin{equation}\label{inequality}
	\max_{L\in D}B_L \le A_2 \le n^{N-2}\max_{L\in D}B_L,
\end{equation}
where we have used the fact that $A_2 = \sum_{L\in D}B_L$. Similarly to \eqref{log A3}, we have
\begin{equation}\label{log BL}
	\begin{split}
		\log B_L =&\;\sum_{i=2}^{N-1}[h(l^i+l^{i-1}+k^+)-h(l^i)-h(l^{i-1}+k^+)]\\
		&\;+\sum_{i=2}^{N-1}[h(m^i+m^{i+1}+k^-)-h(m^i)-h(m^{i+1}+k^-)]+O(\log n)\\
		=&\;\sum_{i=2}^{N-1}n[h(x^i+x^{i-1}+\nu^+)-h(x^i)-h(x^{i-1}+\nu^+)]\\
		&\;+\sum_{i=2}^{N-1}n[h(y^i+y^{i+1}+\nu^-)-h(y^i)-h(y^{i+1}+\nu^-)]+O(\log n),
	\end{split}
\end{equation}
where $x^i = l^i/n$ and $y^i = m^i/n$. For any $\nu\in \mathcal{V}$, we introduce the space
\begin{equation*}
	V(\nu) = \left\{\left(x^{i},y^{i}\right)_{2\le i\le N-1}:\;x^i,y^i\geq0,\;x^{i}+y^{i}=\nu^{i,i+1}\right\},
\end{equation*}
and for any $X=\left(x^{i},y^{i}\right)\in V(\nu)$, we define the function
\begin{equation}\label{formula:F}
	\begin{split}
		F_{\nu}(X)
		=&\sum_{i=2}^{N-1}\left[h\left(x^{i}\right)+h\left(x^{i-1}+\nu^+\right)
		-h\left(x^{i}+x^{i-1}+\nu^+\right)\right] \\
		&+ \sum_{i=2}^{N-1} \left[h\left(y^{i}\right)+h\left(y^{i+1}+\nu^-\right)
		-h\left(y^{i}+y^{i+1}+\nu^-\right)\right],
	\end{split}
\end{equation}
where $x^1=\nu^{12}$ and $y^N=\nu^{N1}$. It then follows from \eqref{inequality} that
\begin{equation}\label{log A2}
	\log A_2 = \max_{L\in D}\log B_L+O(\log n)
	= n\sup_{X\in V(\nu)}F_{\nu}(X)+O(\log n).
\end{equation}
Combining \eqref{joint} and \eqref{LDP}, we obtain
\begin{equation*}\label{log P}
	\begin{split}
		I_J(\nu) &= -\lim_{n\to\infty}\frac{1}{n}\log\mathbb{P}\left(J^c_n=\nu^c,\;\forall c\in\mathcal{C}\right)\\
		&= -\lim_{n\to\infty}\frac{1}{n}\left[\log A_1+\log A_2+\log A_3+\sum_{c\in\mathcal{C}}k^c\log\gamma^c\right].
	\end{split}
\end{equation*}
It then follows from \eqref{log A1}, \eqref{log A3}, and \eqref{log A2} that
\begin{equation}\label{ratefunction}
	\begin{split}
		I_J(\nu) =&\; \left[h\left(\nu^{12}\right)+h\left(\nu^{N1}\right)
		+h\left(\nu^+\right)+h\left(\nu^-\right)-h\left(\nu^{12}+\nu^{N1}+\nu^++\nu^-\right)\right] \\
		&\;+\inf_{X\in V(\nu)}F_{\nu}(X)+\sum_{i\in S}\left[ h\left(\nu_i-\nu^i\right)+h\left(\nu^i\right)
		-h\left(\nu_i\right)\right]-\sum_{c\in\mathcal{C}}\nu^c\log\gamma^c,
	\end{split}
\end{equation}
where $h(x) = x\log x$ and $\nu_i=\sum_{c\ni i}\nu^c$. This gives the expression of the rate function $I_J$ for empirical LE currents. Note that in \eqref{ratefunction}, it is difficult to compute the term $\inf_{X\in V(\nu)} F_{\nu}(X)$. A more explicit expression of this term can be obtained using the Lagrange multiplier method. In Section 2 of Supplementary Material, we have proved that
\begin{equation*}
	\inf_{X\in V(\nu)}F_{\nu}(X) = F_{\nu}(x^i,y^i),
\end{equation*}
where $(x^i,y^i)_{2\leq i\leq N-1}$ is any solution (such solution must exist but may not be unique) of the following set of algebraic equations:
\begin{equation}\label{equation}
	\begin{split}
		\frac{x^{i}}{x^{i-1}+x^{i}+\nu^+}\cdot\frac{x^{i}+\nu^+}{x^{i}+x^{i+1}+\nu^+}
		&= \frac{y^{i}+\nu^-}{y^{i-1}+y^{i}+\nu^-}\cdot\frac{y^{i}}{y^{i}+y^{i+1}+\nu^-},\\
		x^{i} + y^{i} &= \nu^{i,i+1},
	\end{split}
\end{equation}
with $x^1=\nu^{12}$, $x^N=0$, $y^1=0$, and $y^N=\nu^{N1}$.

Thus far, we have assumed that the system starts from state $1$. A natural question is whether the rate function will change when the system starts from other initial distributions. In fact, we can prove that the rate function is independent of the choice of the initial distribution. Note that this is a highly non-trivial result because in the expression \eqref{ratefunction}, the status of state 1 and the status of other states are not equal. The proof is rather complicated and is put in Section 3 of Supplementary Material.

For a general monocyclic system, the expression \eqref{ratefunction} of the rate function is very complicated. This expression can be greatly simplified in two special cases: (i) the case where the system has only three states (any three-state system must be monocyclic) and (ii) the case where the transition from state 1 to state $N$ is forbidden (see Fig. \ref{figure:transitiongraph}(d) for an illustration). For a three-state system, the rate function reduces to (see Appendix~\ref{appendix:threestate} for the proof)
\begin{equation}\label{ratefuntion 3state}
	\begin{split}
		I_J(\nu) =
		\sum_{i\in S} \left[\nu^{i}\log \left(\frac{\nu^{i}/\nu_i}{J^i/J_i}\right) + (\nu_i - \nu^i)\log \left(\frac{(\nu_i - \nu^i)/\nu_i}{(J_i - J^i)/J_i} \right)
		\right]
		+ \sum_{c\in\mathcal{C},|c|\neq 1} \nu^{c} \log \left(\frac{\nu^{c}/\tilde{\nu}}{J^c/\tilde{J}}\right) ,
	\end{split}
\end{equation}
where
\begin{align*}
	\tilde{\nu} &=\sum_{c\in\mathcal{C},|c|\neq 1}\nu^{c}
	= \nu^{12}+\nu^{13}+\nu^{23}+\nu^++\nu^-,\\
	\tilde{J} &=\sum_{c\in\mathcal{C},|c|\neq 1}J^{c}
	= J^{12}+J^{13}+J^{23}+J^++J^-.
\end{align*}
For an $N$-state monocyclic system with the transition from state $1$ to state $N$ being forbidden (Fig. \ref{figure:transitiongraph}(d)), the rate function reduces to (see Appendix~\ref{appendix:threestate} for the proof)
\begin{equation}\label{forbidden}
	I_J(\nu) = \sum_{i\in S}\Bigg[\nu^i\log\left(\frac{\nu^i/\nu_i}{J^i/J_i}\right)
	+\nu^{i,i+1}\log\left(\frac{\nu^{i,i+1}/\nu_i}{J^{i,i+1}/J_i}\right)
	+\left(\nu^{i-1,i}+\nu^+\right)\log\left(\frac{\left(\nu^{i-1,i}+\nu^+\right)/\nu_i}
	{\left(J^{i-1,i}+J^+\right)/J_i}\right)\Bigg].
\end{equation}
Note that the expressions of the rate function in the two special cases are much simpler and more symmetric than the general expression given in \eqref{ratefunction}. Clearly, both expressions have a symmetric form with respect to each state and thus is independent of the choice of the initial distribution. It is well-known that the empirical flows of a Markov chain, i.e. the number of times that each edge is traversed per unit time, also satisfy a large deviation principle and the associated rate function has the form of relative entropy (see Section \ref{FDTST} for details). Interestingly, we find that the rate functions given in \eqref{ratefuntion 3state} and \eqref{forbidden} also have a functional form similar to relative entropy.

The large deviations for empirical LE currents $(J^c_n)_{c\in\mathcal{C}}$ can be directly applied to establish the large deviations for empirical net LE currents $(\tilde{J}^c_n)_{c\in\mathcal{C}}$. Since the empirical net LE currents vanish for any one-state and two-state cycles and since $\tilde{J}^+_n = -\tilde{J}^-_n$ for the two $N$-state cycles $(1,2,\cdots,N)$ and $(1,N,\cdots,2)$, we only need to focus on the empirical net currents $\tilde{J}^+_n$ of cycle $(1,2,\cdots,N)$. By the contraction principle, we have
\begin{equation}\label{tilde I J}
	\begin{split}
		\mathbb{P}\left(\tilde{J}^{+}_n = x\right)
		&=\;\mathbb{P}\left(J^{+}_n-J^{-}_n = x\right)\\
		&=\;\sum_{\nu^{+}-\nu^{-}=x}\mathbb{P}\left(J^{c}_n=\nu^{c},\forall c\in\mathcal{C}\right)\\
		&\propto\sum_{\nu^{+}-\nu^{-}=x} e^{-nI_J(\nu)},\;\;\;n\to\infty.
	\end{split}
\end{equation}
This shows that the empirical net LE current $\tilde{J}^+_n$ satisfies a large deviation principle with rate function
\begin{equation}\label{ratefuntion net LE}
	I_{\tilde{J}}(x)=\inf_{\{\nu\in\mathcal{V}:\;\nu^{+}-\nu^{-}= x\}}I_J(\nu).
\end{equation}

\subsection{Large deviations for ST currents of general Markov chains}\label{FDTST}
We next focus on the large deviations for empirical ST currents of a general Markov chain. In fact, the large deviations for empirical net ST currents have been investigated and the symmetry of the rate function has been obtained in \cite{bertini2015flows}. Here we focus on the large deviations for empirical (absolute) ST currents. To this end, we first recall the large deviations for empirical flows \cite{den2008large}.

Recall that the empirical flow of edge $\langle i,j\rangle$ up to time $n$ is defined as
\begin{equation*}
	R_n(i,j) = \frac{1}{n}\sum_{m=1}^n1_{\{\xi_{m-1}=i,\xi_m=j\}}.
\end{equation*}
Intuitively, $R_n(i,j)$ represents the number of times that edge $\langle i,j\rangle$ is traversed per unit time. Note that under the periodic boundary condition, the empirical flows $(R_n(i,j))_{\langle i,j\rangle\in E}$ must lie in the space
\begin{equation*}
	\mathcal{M} = \bigg\{(R_n(i,j))_{\langle i,j\rangle\in E}:\;R_n(i,j)\geq 0,\;\sum_{i,j\in S}R(i,j) = 1,\;
	\sum_{j\in S}R(i,j)=\sum_{j\in S}R(j,i)\bigg\}.
\end{equation*}
It is well known that the empirical flows $(R_n(i,j))_{\langle i,j\rangle\in E}$ satisfy the following large deviation principle:
\begin{equation*}
	\mathbb{P}(R_n(i,j)=R(i,j),\;\forall\langle i,j\rangle\in E)\propto e^{-nI_{\mathrm{flow}}(R)},\;\;\;n\to\infty,
\end{equation*}
where the rate function $I_{\mathrm{flow}}:\mathcal{M}\rightarrow[0,\infty]$ is given by
\begin{equation*}\label{I pair}
	I_{\mathrm{flow}}(R) = \sum_{\langle i,j\rangle\in E}R(i,j)\log\frac{R(i,j)}{R(i)p_{ij}},
\end{equation*}
with $R(i)=\sum_{j\in S}R(i,j)$. Clearly, the rate function for empirical flows has the form of relative entropy. For any chord $l$ of a fixed spanning tree $T$, let $H^{c_l}$ be a function on $E$ defined by
\begin{equation}\label{cycle function2}
	H^{c_l}(i,j)
	=\left\{\begin{aligned}
		1, &   && \text{if (i) }\langle i,j\rangle \in T\text{ and }\langle i,j\rangle \in c_l\text{ or (ii) } \langle i,j\rangle=l,\\
		-1,&   && \text{if }\langle i,j\rangle \in T, \langle i,j\rangle\notin c_l,\text{ and }\langle j,i\rangle \in c_l,\\
		0, &   && \text{otherwise}.\\
	\end{aligned}\right.
\end{equation}
In fact, the empirical flow $R_n(i,j)$ can be represented as the weighted sum of $H^{c_l}(i,j)$ with the weights being all empirical ST currents \cite{kalpazidou2007cycle}, i.e.
\begin{equation*}
	R_n(i,j) = \sum_{c_l\in\mathcal{L}}Q^{c_l}_nH^{c_l}(i,j),\;\;\;\langle i,j\rangle\in E.
\end{equation*}
It was further proved in \cite{kalpazidou2007cycle} that this representation is unique. In other words, if $R_n = \sum_{c_l\in\mathcal{L}}\mu^{c_l}H^{c_l}$ for some coefficients $\mu^{c_l}$, then we must have $\mu^{c_l} = Q^{c_l}_n$ for any $c_l\in\mathcal{L}$. It then follows from the uniqueness of the above representation that
\begin{equation*}
	\begin{split}
		\mathbb{P}(Q_n^{c_l}=\mu^{c_l},\;\forall c_l\in\mathcal{L})
		&=\mathbb{P}\bigg(R_n(i,j)=\sum_{c_l\in\mathcal{L}}\mu^{c_l}H^{c_l}(i,j),\;\forall\langle i,j\rangle\in E\bigg)\\
		&\propto e^{-n I_{\mathrm{flow}}\left(\sum_{c_l\in\mathcal{L}}\mu^{c_l}H^{c_l}\right)},\;\;\;n\rightarrow\infty.
	\end{split}
\end{equation*}
This shows that the empirical ST currents $(Q_n^{c_l})_{c_l\in\mathcal{L}}$ satisfy a large deviation principle with rate function
\begin{equation}\label{formula:I_Q}
	I_Q(\mu)=I_{\mathrm{flow}}\left(\sum_{c_l\in\mathcal{L}}\mu^{c_l}H^{c_l}\right).
\end{equation}

Thus far, we have obtained the explicit expressions of the rate function for empirical LE currents of a monocyclic system and the rate function for empirical ST currents of a general system. A natural question is what is the relationship between the two rate functions. To see this, recall that ST currents can be represented by LE currents as $Q_n^{c_l} = \sum_{c\ni l}J^c_n$. It thus follows from the contraction principle that
\begin{align*}
	\mathbb{P}\left(Q_n^{c_l}=\mu^{c_l},\;\forall l\in\mathcal{L}\right)
	&= \mathbb{P}\left(\sum_{c\ni l}J^c_n=\mu^{c_l},\;\forall l\in\mathcal{L}\right)\\
	&= \sum_{\sum_{c\ni l}\nu^c=\mu^{c_l}}
	\mathbb{P}\left(J^c_n=\nu^c,\;\forall c\in\mathcal{C}\right)\\
	&\propto \sum_{\sum_{c\ni l}\nu^c=\mu^{c_l}}e^{-nI_J(\nu)},
	\;\;\;n\rightarrow\infty.
\end{align*}
This shows that the rate functions for empirical LE and ST currents are connected by
\begin{equation*}
	I_Q(\mu) = \inf_{\{\nu\in\mathcal{V}:\;\sum_{c\ni l}\nu^c=\mu^{c_l}\}}I_J(\nu).
\end{equation*}
It is straightforward to prove that the rate function $I_Q$ given above coincides with the one given in  \eqref{formula:I_Q} for monocyclic systems.

The large deviations for empirical ST currents $(Q^{c_l}_n)_{c_l\in\mathcal{L}}$ can also be used to establish the large deviations for empirical net ST currents $(\tilde{Q}^{c_l}_n)_{c_l\in\mathcal{L}}$. Since the empirical net ST currents vanish for all one-state and two-state cycles, we only need to focus on cycles with three or more states. Let $c_{l_1},c_{l_2},\cdots,c_{l_s}$ be all cycles with three or more states in the fundamental set so that any two of them are not reversed cycles of each other. By the contraction principle, the empirical net ST currents $(\tilde{Q}^{c_{l_i}})_{1\le i\le s}$ of these cycles satisfy a large deviation principle with rate function
\begin{equation}\label{I_Q2}
	I_{\tilde{Q}}(x)=\inf_{\{\mu\in\mathcal{M}:\;\mu^{c_{l_{i}}}-\mu^{c_{l_{i}}-}= x_i,\;\forall 1\le i\le s\}}I_Q(\mu).
\end{equation}

\subsection{Applications in single-molecule enzyme kinetics}\label{sec:example}
\begin{figure}[h]
		\centering\includegraphics[width=1.0\textwidth]{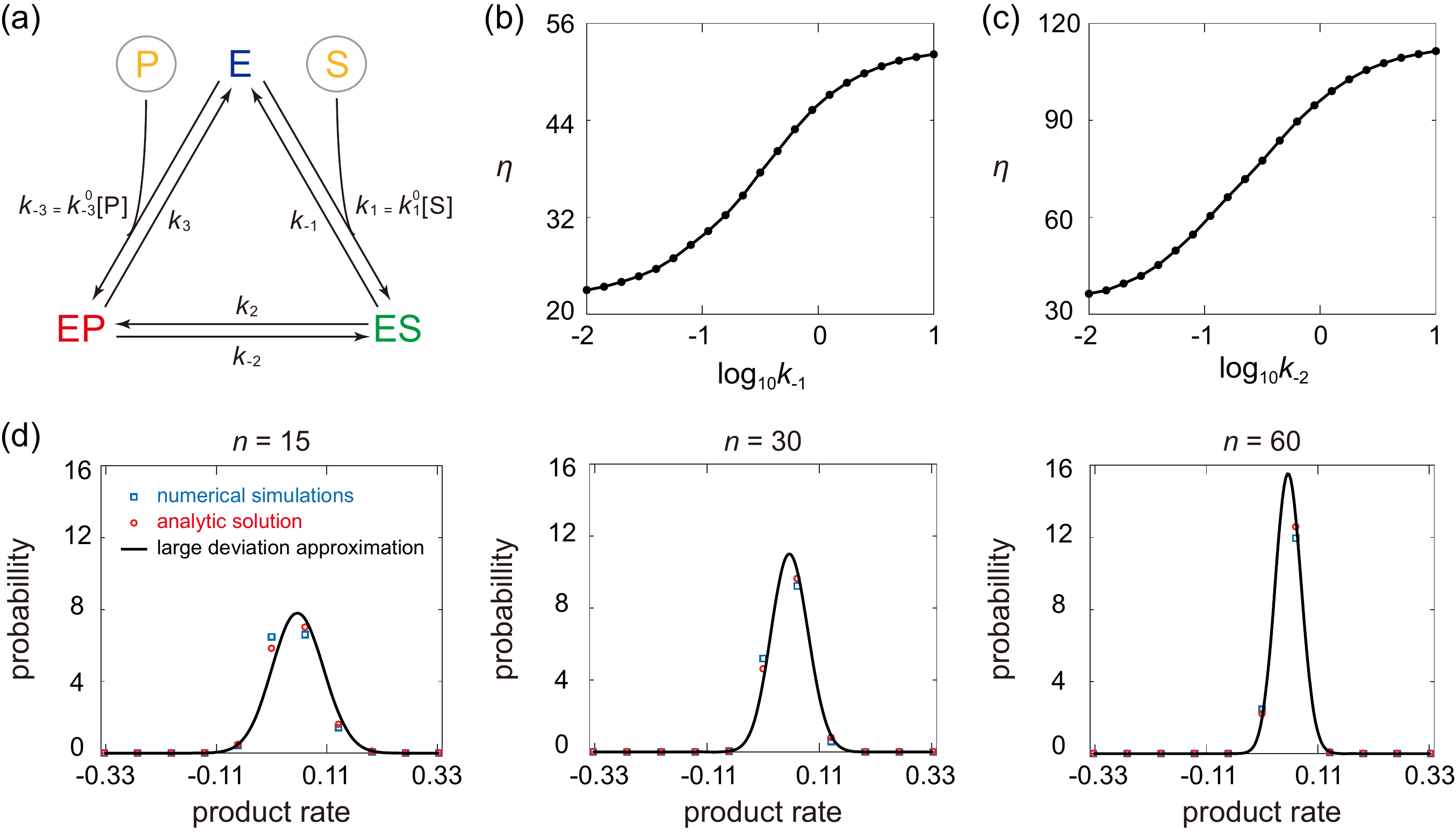}
		\caption{\textbf{Three-step mechanism of a reversible enzymatic reaction.} \textbf{(a)} Kinetic scheme of a three-step reversible enzyme reaction. Here $k^0_1$ and $k^0_{-3}$ are second-order rate constants, and $k_1 = k^0_1[S]$ and $k_{-3} = k^0_{-3}[P]$ are pseudo-first-order rate constants. From the perspective of a single enzyme molecule, the reaction is unimolecular and cyclic. \textbf{(b)} Noise $\eta$ in the product rate versus the rate constants $k_{-1}$. The parameters are chosen as $\tau=0.01$, $n=15$, $k^0_1=2k_{-1}$, $k_2=1$, $k_{-2}=1$, $k_3=1$, $k^0_{-3}=0.1$, $[P]=1$, and $[S]$ is tuned so that $\langle\tilde{J}^+_n\rangle$ remains invariant. \textbf{(c)} Noise $\eta$ in the product rate versus the rate constants $k_{-2}$. The parameters are tuned as $\tau=0.01$, $n=15$, $k^0_1=1.2$, $k_{-1}=0.6$, $k_2=k_{-2}$, $k_3=1$, $k^0_{-3}=0.1$, $[P]=1$, and $[S]$ is tuned so that $\langle\tilde{J}^+_n\rangle$ remains invariant. \textbf{(d)} Distribution of the product rate $\tilde{J}^+_n$ as time $n$ increases. The blue squares are the ones obtained using stochastic simulations, the red circles are the ones obtained using the exact joint distribution \eqref{joint}, and the black curves are the ones obtained using the exact rate function \eqref{ratefuntion 3state} and large deviation approximation \eqref{LDP}. The parameters are chosen as $\tau=0.1$, $k^0_1=4$, $k_{-1}=2$, $k_2=5$, $k_{-2}=1$, $k_3=6$, $k^0_{-3}=0.1$, and $[S]=[P]=1$.}\label{figure:example}
	\end{figure}

	As an application of our theoretical results, we consider the following three-step mechanism of a reversible enzymatic reaction \cite{ge2008waiting, ge2012stochastic}:
	\begin{equation*}\label{eq:model}
		E+S \underset{k_{-1}}{\stackrel{k_{1}^{0}}{\rightleftharpoons}} ES \underset{k_{-2}}{\stackrel{k_{2}}{\rightleftharpoons}} EP \underset{k_{-3}^{0}}{\stackrel{k_{3}}{\rightleftharpoons}} E+P,
	\end{equation*}
	where $E$ is an enzyme turning the substrate $S$ into the product $P$. If there is only one enzyme molecule, then it may convert stochastically among three conformal states: the free enzyme $E$, the enzyme-substrate complex $ES$, and the enzyme-product complex $EP$. For simplicity, we assume that the enzyme reaction is in an open system with the concentrations of $S$ and $P$ sustained by an external agent \cite{qian2007phosphorylation}. Then from the enzyme perspective, the kinetics is stochastic and cyclic with pseudo-first-order rate constants $k_1 = k^0_1[S]$ and $k_{-3} = k^0_{-3}[P]$, where $[S]$ and $[P]$ are the sustained concentrations of $S$ and $P$, respectively (Fig. \ref{figure:example}(a)). Note that the time variable of the enzyme reaction is continuous. However, in experiments, we are only able to observe the system at multiple discrete time points. If we record the conformal state of the enzyme molecule at a series of time points with interval $\tau$, then the system can be modelled as a three-state discrete-time Markov chain, which coincides with the model studied in this paper. Let $Q = (q_{ij})$ be the transition rate matrix of the continuous-time system shown in Fig. \ref{figure:example}(a). Then the transition probability matrix of the discrete-time system is given by $P = (p_{ij}) = e^{\tau Q}$ \cite{norris1998markov}. The discrete-time system serves as a good approximation of the continuous-time system when the interval $\tau$ is small.
	
	Note that for the cyclic kinetics illustrated in Fig. \ref{figure:example}(a), a substrate molecule $S$ is converted into a product molecule $P$ whenever the clockwise cycle $C^+ = (E,ES,EP)$ is formed, and a product molecule $P$ is converted into a substrate molecule $S$ whenever the counterclockwise cycle $C^- = (E,EP,ES)$ is formed. Thus the rate of product formation, also called \emph{product rate}, of the enzyme reaction, i.e. the net conversion of $S$ into $P$ per unit time, is exactly the net LE current $\tilde{J}^+_n = J^+_n-J^-_n$. Previous studies \cite{ge2012stochastic} mainly focus on the long-time mean product rate
	\begin{equation*}
		\lim_{n\to\infty}\tilde{J}^+_n = \tilde{J}^+ = \frac{\gamma^+-\gamma^-}{C},
	\end{equation*}
	where $\gamma^+ = p_{12}p_{23}p_{31}$, $\gamma^- = p_{13}p_{32}p_{21}$, and
	\begin{equation*}
		C = \sum_{i=1}^3[(1-p_{i-1,i-1})(1-p_{i+1,i+1})-p_{i-1,i+1}p_{i+1,i-1}].
	\end{equation*}
	The analytical results derived in previous sections allow us to investigate the finite-time fluctuation behavior of the product rate $\tilde{J}^+_n$. In experiments, the size of fluctuations, also called \emph{noise}, in the product rate is often measured by the coefficient of variation $\eta = \sigma/\mu$, where $\mu = \langle\tilde{J}^+_n\rangle$ is the mean and $\sigma$ is the standard deviation \cite{paulsson2005models}. Note that we have obtained the exact joint distribution of empirical LE currents in Section \ref{sec:joint}. Using the joint distribution, it is easy to calculate all moments, including the mean and standard deviation, of the product rate $\tilde{J}^+_n$.
	
	In Fig. \ref{figure:example}(b),(c), we illustrate noise $\eta$ as a function of the rate constants $k_{-1}$ and $k_{-2}$. Here $k_{-1}$ and $k_{-2}$ are varied while keeping $k^0_1/k_{-1}$ and $k_2/k_{-2}$ as constant, and the substrate concentration $[S]$ is tuned so that the mean product rate $\mu$ remains invariant (examining protein noise while fixing the protein mean is a common strategy in molecular biology experiments \cite{schmiedel2015microrna}). From Fig. \ref{figure:example}(b), we see that noise in the product rate becomes larger as $k^0_1$ and $k_{-1}$ increase (while keeping their ratio as constant). Note that when $k^0_1$ and $k_{-1}$ are both large, the reaction $E+S\rightleftharpoons ES$ will reach rapid pre-equilibrium and this is widely known as rapid equilibrium assumption in enzyme kinetics \cite{beard2008chemical}. Our results show that rapid equilibrium between the enzyme states $E$ and $ES$ leads to large fluctuations in the product rate. Similarly, from Fig. \ref{figure:example}(c), we find that noise in the product rate also becomes larger as $k_2$ and $k_{-2}$ increase (while keeping their ratio as constant). Note that when $k_2$ and $k_{-2}$ are both large, the two enzyme states $ES$ and $EP$ will reach rapid pre-equilibrium and thus can be combined into a single state \cite{bo2016multiple, jia2016model}. In this case, the three-step enzyme reaction reduces to the classical two-step Michaelis-Menten enzyme kinetics
	\begin{equation*}
		E+S \rightleftharpoons ES\rightleftharpoons E+P,
	\end{equation*}
	This implies that compared to the two-step Michaelis-Menten kinetics, the three-step kinetics results in smaller fluctuations in the product rate.
	
	While the exact joint distribution for LE currents derived in Section \ref{LELDP} can be used to study the fluctuations in the product rate, it is computationally very slow because we need to calculate a large number of factorials and combinatorial numbers (see \eqref{joint}), especially when time $n$ and the number of states $N$ are large. Fortunately, the large deviations for LE currents studied in Section \ref{sec:LDP} can be used to provide a much more efficient computational method of the joint distribution. Specifically, we only need to compute the rate function $I_J(\nu)$ using \eqref{ratefuntion 3state} and then apply \eqref{LDP} to construct an approximation of the joint distribution. In Fig. \ref{figure:example}(d), we compare the distribution of the production rate $\tilde{J}^+_n$ obtained by using stochastic simulations (blue squares), the analytical solution (red circles), and the large deviation approximation (black curves). As expected, the analytical solution coincides perfectly with stochastic simulations. Interestingly, we find that the approximate distribution obtained based on the large deviation theory is in good agreement with the analytical solution when $n\geq 15$ and they become practically indistinguishable when $n\geq 30$. According to our simulations, when $n=30$, compared with the analytical solution, the large deviation approximation can save the computational time by over $99\%$. This suggests that the large deviation principle studied in this paper is very useful because it enables a fast exploration of large swaths of parameter space.

\section{Fluctuation theorems for cycle currents}\label{sec:FT}
Next we investigate the fluctuation relations satisfied by the two types of cycle currents. In Section \ref{sec:FT monocyclic}, using trajectory reversal method, we obtain a symmetric relation for LE currents of a monocyclic system that is even stronger than the classical transient and integral fluctuation theorems. In Section \ref{sec:FT general}, we generalize the fluctuation relations to a general system, and reveal their connection with the second law of thermodynamics. In Section \ref{sec:FT ST}, we explore the fluctuation relations for ST currents of a general system and compare them with the fluctuation relations for LE currents.

\subsection{Fluctuation theorems for LE currents of monocyclic Markov chains}\label{sec:FT monocyclic}
An important question is whether empirical cycle currents satisfy various fluctuation theorems. In fact, the transient fluctuation theorem for net LE currents has been investigated in \cite{qian2006generalized, andrieux2007network}. Here we will prove a symmetric relation for a monocyclic system that is even stronger than the transient fluctuation theorem. For convenience, we write the two $N$-state cycles of a monocyclic system as $C^+ = (1,2,\cdots,N)$ and $C^- = (1,N,\cdots,2)$. Let $N^+_n$ and $N^-_n$ denote the number of times that cycles $C^+$ and $C^-$ are formed up to time $n$, respectively. The strong symmetric relation for LE currents is given by
\begin{equation}\label{sm}
	\begin{split}
		&\;k^+\mathbb{P}\left(N^+_n=k^+,N^-_n=k^--1,N^c_n=k^c,\;\forall c\neq C^+,C^-\right)\\
		=&\; \left(\frac{\gamma^+}{\gamma^-}\right)k^-\mathbb{P}\left(N^+_n=k^+-1,N^-_n=k^-,N^c_n=k^c,\;\forall c\neq C^+,C^-\right),
	\end{split}
\end{equation}
where $\gamma^+ = p_{12}p_{23}\cdots p_{N1}$ and $\gamma^- = p_{21}p_{32}\cdots p_{N1}$ are the product of transition probabilities along cycles $C^+$ and $C^-$, respectively. In fact, a similar equality has been obtained recently for another type of cycle currents defined in the sequence matching manner \cite{patrick2021cycle}. We next give the proof of \eqref{sm} for LE currents. Under the periodic boundary condition, it follows from \eqref{joint} that
\begin{align*}
	&\;\mathbb{P}\left(N^+_n=k^+,N^-_n=k^--1,N^c_n=k^c,\;\forall c\neq C^+,C^-\right)\\
	=&\; (\gamma^+)^{k^+}(\gamma^-)^{k^--1}\prod_{c\neq C^+,C^-}
	\left(\gamma^c\right)^{k^c}\left|G_n(k^+,k^--1,(k^c)_{c\neq C^+,C^-})\right|,
\end{align*}
where $G_n(k^+,k^--1,(k^c)_{c\neq C^+,C^-})$ is the collection of all possible trajectories up to time $n$ so that cycle $C^+$ is formed $k^+$ times, cycle $C^-$ is formed $k^--1$ times, and any other cycle $c\neq C^+,C^-$ is formed $k^c$ times. For simplicity of notation, we rewrite the above equation as
\begin{equation*}
	\mathbb{P}\left(N^+_n=k^+,N^-_n=k^--1,\cdots\right)
	= (\gamma^+)^{k^+}(\gamma^-)^{k^--1}\prod_{c\neq C^+,C^-}\left(\gamma^c\right)^{k^c}|G_n(k^+,k^--1,\cdots)|.
\end{equation*}
Similarly, replacing $k^+$ by $k^+-1$ and replacing $k^--1$ by $k^-$ in the above equation, we obtain
\begin{equation*}
	\mathbb{P}\left(N^+_n=k^+-1,N^-_n=k^-,\cdots\right)
	= (\gamma^+)^{k^+-1}(\gamma^-)^{k^-}\prod_{c\neq C^+,C^-}\left(\gamma^c\right)^{k^c}|G_n(k^+-1,k^-,\cdots)|.
\end{equation*}
Hence, to prove \eqref{sm}, we only need to show that
\begin{equation}\label{equal}
	k^+|G_n(k^+,k^--1,\cdots)| = k^-|G_n(k^+-1,k^-,\cdots)|.
\end{equation}
For any trajectory $\{\xi_0,\xi_1,\cdots,\xi_n\}$ lying in $G_n(k^+,k^--1,\cdots)$, since cycle $C^+$ is formed $k^+$ times, there are $k^+$\\ beginning times (the times that $C^+$ begins to form) and $k^+$ ending times (the times that $C^+$ has been formed) for this cycle. Let $T^{\mathrm{begin}}_i$ and $T^{\mathrm{end}}_i$ denote the $i$th beginning and ending times for cycle $C^+$, respectively. For example, for the trajectory given in Table \ref{trajectory}, the first beginning time for cycle $c= (1,2,3,4)$ is $n = 0$ and the first ending time is $n = 7$. If we reverse the trajectory $\{\xi_0,\xi_1,\cdots,\xi_n\}$ between $T^{\mathrm{begin}}_i$ and $T^{\mathrm{end}}_i$, then we obtain a new trajectory  $\{\tilde\xi_0,\tilde\xi_1,\cdots,\tilde\xi_n\}$, which is given by
\begin{equation*}
	\tilde{\xi}_m =\left\{\begin{aligned}
		&\xi_{T^{\mathrm{begin}}_i+T^{\mathrm{end}}_i-m}, &&
		\text{if } T^{\mathrm{begin}}_i\le m\le T^{\mathrm{end}}_i\,\\
		&\xi_m,    && \text{otherwise}.\\
	\end{aligned}\right.
\end{equation*}
Clearly, the reversed trajectory must lie in $G_n(k^+-1,k^-,\cdots)$. Since cycle $C^+$ is formed $k^+$ times, there are $k^+|G_n(k^+,k^--1,\cdots)|$ possible reversed trajectories. Among these reversed trajectories, $k^-$ trajectories are exactly the same and are counted repetitively. For example, if $C^+ = (1,2,3)$ and $C^- = (1,3,2)$, then the trajectories $\{1,2,3,1,2,3,1,3,2\}$ and $\{1,2,3,1,3,2,1,2,3\}$ in $G_8(2,1,\cdots)$ can both be reversed to the trajectory $\{1,2,3,1,3,2,1,3,2\}$ in $G_8(1,2,\cdots)$, and thus are counted twice. As a result, the number of possible trajectories in $G_n(k^+-1,k^-,\cdots)$ is given by
\begin{equation*}
	|G_n(k^+-1,k^-,\cdots)| = \frac{k^+}{k^-}|G_n(k^+,k^--1,\cdots)|,
\end{equation*}
which is exactly \eqref{equal}. Thus we have proved the strong symmetric relation given by \eqref{sm}. Applying the symmetric relation $|k^+-k^-|$ times, we obtain the transient fluctuation theorem for LE currents
\begin{equation}\label{tft}
	\mathbb{P}\left(N^+_n=k^+,N^-_n=k^-,\cdots\right)
	= \mathbb{P}\left(N^+_n=k^-,N^-_n=k^+,\cdots\right)\left(\frac{\gamma^+}{\gamma^-}\right)^{k^+-k^-}.
\end{equation}
Thus far, we have proved the symmetric relation \eqref{sm} and transient fluctuation theorem \eqref{tft} under the periodic boundary condition. Without the periodic boundary condition, these two equalities are also valid for monocyclic systems; the proof is similar and thus is omitted.

The transient fluctuation theorem can be used to prove other two types of fluctuation theorems. To see this, recall that the moment generating function of empirical LE currents is defined as
\begin{equation*}
	g_n(\lambda^+,\lambda^-,\cdots)
	= \left\langle e^{\lambda^+N^+_n+\lambda^-N^-_n+\sum_{c\neq C^+,C^-}\lambda^cN^c_n}\right\rangle,
\end{equation*}
where $\langle A\rangle$ denotes the mean of $A$. Then the following Kurchan-Lebowitz-Spohn-type fluctuation theorem holds:
\begin{align*}
	g_n(\lambda^+,\lambda^-,\cdots)
	&= \sum_{k}e^{\sum_{c\in\mathcal{C}}\lambda^ck^c}\mathbb{P}\left(N^+=k^+,N^-=k^-,\cdots\right)\\
	&= \sum_{k}e^{\sum_{c\in\mathcal{C}}\lambda^ck^c}
	\mathbb{P}\left(N^+=k^-,N^-=k^+,\cdots\right)\left(\frac{\gamma^+}{\gamma^-}\right)^{k^+-k^-}\\
	&= \sum_{k}e^{\cdots+\left(\lambda^+-\log\frac{\gamma^+}{\gamma^-}\right)k^++
		\left(\lambda^--\log\frac{\gamma^-}{\gamma^+}\right)k^-}\mathbb{P}(N^+=k^-,N^-=k^+,\cdots)\\
	&= \left\langle e^{\left(\lambda^--\log\frac{\gamma^+}{\gamma^-}\right)N^+_n+
		\left(\lambda^++\log\frac{\gamma^+}{\gamma^-}\right)N^-_n+\cdots}\right\rangle\\
	&= g_n\left(\lambda^--\log\frac{\gamma^+}{\gamma^-},
	\lambda^++\log\frac{\gamma^+}{\gamma^-},\cdots\right),
\end{align*}
where $\log(\gamma^+/\gamma^-)$ is the affinity of cycle $C^+$ \cite{seifert2012stochastic}. We next consider the long-time limit behavior of a monocyclic system. As $n\rightarrow\infty$, it is easy to see that
\begin{align*}
	e^{-nI_J(\nu^+,\nu^-,\cdots)}
	&\propto \mathbb{P}\left(J^+_n=\nu^+,J^-_n=\nu^-,\cdots\right)\\
	&= \mathbb{P}\left(J^+_n=\nu^-,J^-_n=\nu^+,\cdots\right)
	\left(\frac{\gamma^+}{\gamma^-}\right)^{n(\nu^+-\nu^-)}\\
	&\propto e^{-n\left[I_J(\nu^-,\nu^+,\cdots)-\left(\log\frac{\gamma^+}{\gamma^-}\right)
		(\nu^+-\nu^-)\right]}.
\end{align*}
This yields the Gallavotti-Cohen-type fluctuation theorem
\begin{equation}\label{G-C type fluctuation}
	I_J(\nu^+,\nu^-,\cdots)=I_J(\nu^-,\nu^+,\cdots)-\left(\log\frac{\gamma^+}{\gamma^-}\right)(\nu^+-\nu^-).
\end{equation}

Similarly, we can also obtain the fluctuation theorems for net LE currents. For a monocyclic system, we only need to focus on the empirical net LE current $\tilde{J}^+_n$ of cycle $C^+$. Let $\tilde{g}_n(\lambda) = \langle e^{\lambda n\tilde{J}^+_n}\rangle$ be the moment generating function of $\tilde{J}^+_n$ and let $I_{\tilde{J}}(x)$ be the rate function of $\tilde{J}^+_n$ given in \eqref{ratefuntion net LE}. The various fluctuation theorems for net LE currents were first obtained in \cite{qian2006generalized} and are summarized as follows. The proof is similar and thus is omitted.

1) Transient fluctuation theorem:
\begin{equation*}
	\frac{\mathbb{P}(\tilde{J}^+_n=x)}{\mathbb{P}(\tilde{J}^+_n=-x)}
	= \left(\frac{\gamma^+}{\gamma^-}\right)^{nx}.
\end{equation*}

2) Kurchan-Lebowitz-Spohn-type fluctuation theorem:
\begin{equation*}
	\tilde{g}_n(\lambda)=\tilde{g}_n\left(-\left(\lambda+\log\frac{\gamma^+}{\gamma^-}\right)\right).
\end{equation*}

3) Integral fluctuation theorem: Taking $\lambda = -\log (\gamma^+/\gamma^-)$ in the above equation yields
\begin{equation*}
	\left\langle e^{\lambda n\tilde{J}^+_n}\right\rangle=1.
\end{equation*}

4) Gallavotti-Cohen-type fluctuation theorem:
\begin{equation*}
	I_{\tilde{J}}(x)=I_{\tilde{J}}(-x)-\left(\log\frac{\gamma^+}{\gamma^-}\right)x.
\end{equation*}

\subsection{Fluctuation theorems for LE currents of general Markov chains}\label{sec:FT general}
We have seen that various symmetric relations and fluctuation theorems hold for LE currents of a monocyclic system. A natural question is whether these results can be extended to a general Markov chain. Before stating the results, we recall the definition of similar cycles \cite{jia2016cycle}. Let $c_1=(i_1,i_2,\cdots,i_s)$ and $c_2=(j_1,j_2,\cdots,j_r)$ be two cycles. Then $c_1$ and $c_2$ are called \emph{similar} if $s=r$ and $\{i_1,i_2,\cdots,i_s\}=\{j_1,j_2,\cdots,j_r\}$. In other words, two cycles are similar if they pass through the same set of states. For example, the following six cycles:
\begin{equation*}
	(1,2,3,4),\;(1,2,4,3),\;(1,3,2,4),\;(1,3,4,2),\;(1,4,2,3),\;(1,4,3,2)
\end{equation*}
are similar. Note that any cycle $C$ and its reversed cycle $C-$ must be similar.

We first focus on empirical LE currents $(J^c_n)_{c\in\mathcal{C}}$, where $J^c_n = N^c_n/n$. For a general Markov chain, if cycles $c_1$ and $c_2$ are similar, then the following symmetric relation holds:
\begin{equation}\label{generalized transient}
	\frac{k^{c_1}\Pnum(N^{c_1}_n=k^{c_1},N^{c_2}_n=k^{c_2}-1,N^{c}_n=k^{c},\;\forall c\neq c_1,c_2)}
	{k^{c_2}\Pnum(N^{c_1}_n=k^{c_1}-1,N^{c_2}_n=k^{c_2},N^{c}_n=k^{c},\;\forall c\neq c_1,c_2)}
	= \frac{\gamma^{c_1}}{\gamma^{c_2}}.
\end{equation}
If we choose $c_1$ and $c_2$ to be some cycle $C^+$ and its revered cycle $C^-$, then this equality reduces to
\begin{equation*}
	\frac{k^+\Pnum(N^+_n=k^+,N^-_n=k^--1,N^c_n=k^c,\;\forall c\neq C^+,C^-)}
	{k^-\Pnum(N^+_n=k^+-1,N^-_n=k^-,N^c_n=k^c,\;\forall c\neq C^+,C^-)}
	= \frac{\gamma^+}{\gamma^-}.
\end{equation*}
This can be viewed as a generalization of \eqref{sm} in the monocyclic case. Applying \eqref{generalized transient} repeatedly gives the following transient fluctuation theorem for LE currents:
\begin{equation}\label{transient}
	\frac{\Pnum(N^{c_1}_n=k^{c_1},N^{c_2}_n=k^{c_2},N^{c}_n=k^{c},\;\forall c\neq c_1,c_2)}
	{\Pnum(N^{c_1}_n=k^{c_2},N^{c_2}_n=k^{c_1},N^{c}_n=k^{c},\;\forall c\neq c_1,c_2)}
	= \left(\frac{\gamma^{c_1}}{\gamma^{c_2}}\right)^{k^{c_1}-k^{c_2}}.
\end{equation}
This shows that if the cycles $c_1$ and $c_2$ are similar, then the joint distribution of empirical LE currents satisfies a symmetric relation under the exchange of $k^{c_1}$ and $k^{c_2}$. Actually, the proof of \eqref{transient} has been given in \cite{jia2016cycle} under the restrictions that all cycles under consideration pass through a common state $i\in S$ and the Markov chain also starts from state $i$. Fortunately, this technical assumption can be removed and the result holds generally (manuscript in preparation).

We next consider empirical net LE currents $(\tilde{J}^c_n)_{c\in\mathcal{C}}$. Let $c_1,c_2,\cdots,c_r$ be all cycles with three or more states in the cycle space so that any two of them are not reversed cycles of each other (the empirical net LE currents for one-state and two-state cycles vanish and do not need to be considered). It then follows from \eqref{transient} that
\begin{equation*}
	{\small\begin{split}
			&\; \Pnum\left(\tilde{J}^{c_1}_n=x_1,\tilde{J}^{c_m}_n=x_m,\;\forall 2\le m\le r\right) \\
			=&\; \Pnum\left(N^{c_1}_n-N^{c_1-}_n=nx_1,N^{c_m}_n-N^{c_m-}_n=nx_m,\;\forall 2\le m\le r\right) \\
			=&\; \sum_{k^{c_i}-k^{c_i-}=nx_i,\;\forall 1\le i\le r }\Pnum\left(N^{c_1}_n=k^{c_1},N^{c_1-}_n=k^{c_1-},  N^{c_m}_n=k^{c_m},N^{c_m-}_n=k^{c_m-},\;\forall 2\le m\le r\right) \\
			=&\; \sum_{k^{c_i}-k^{c_i-}=nx_i,\;\forall  1\le i\le r }\Pnum\left(N^{c_1}_n=k^{c_1-},N^{c_1-}_n=k^{c_1}, N^{c_m}_n=k^{c_m},N^{c_m-}_n=k^{c_m-},\;\forall 2\le m\le r\right)\left(\frac{\gamma^{c_1}}{\gamma^{c_1-}}\right)^{nx_1} \\
			=&\; \Pnum\left(N^{c_1}_n-N^{c_1-}_n=-nx_1,N^{c_m}_n-N^{c_m-}_n=nx_m,\;\forall 2\le m\le r\right)e^{nx_1\log\frac{\gamma^{c_1}}{\gamma^{c_1-}}} \\
			=&\; \Pnum\left(\tilde{J}^{c_1}_n=-x_1,\tilde{J}^{c_m}_n=x_m,\;\forall 2\le m\le r\right)e^{nx_1\log\frac{\gamma^{c_1}}{\gamma^{c_1-}}}.
	\end{split}}
\end{equation*}
Hence we have obtained the following transient fluctuation theorems for net LE currents:
\begin{equation}\label{strong}
	\frac{\Pnum(\tilde{J}^{c_1}_n=x_1,\tilde{J}^{c_m}_n=x_m,\;\forall 2\le m\le r)}
	{\Pnum(\tilde{J}^{c_1}_n=-x_1,\tilde{J}^{c_m}_n=x_m,\;\forall 2\le m\le r)}
	= e^{nx_1\log\frac{\gamma^{c_1}}{\gamma^{c_1-}}}.
\end{equation}
This shows that the joint distribution of empirical net LE currents satisfies a symmetric relation when any $x_i$ is replaced by $-x_i$. In fact, this result which was first found in \cite{qian2006generalized} for a monocyclic system and further generalized in \cite{andrieux2007network} to a general system, while the proof is not totally rigorous. If we change $x_i$ to $-x_i$ one by one for $1\leq i\leq r$ in the above equation, then we obtain
\begin{equation}\label{weak}
	\begin{split}
		\frac{\Pnum\left(\tilde{J}^{c_1}_n=x_1,\tilde{J}^{c_2}_n=x_2,\cdots,\tilde{J}^{c_{r}}_n=x_{r}\right)}
		{\Pnum\left(\tilde{J}^{c_1}_n=-x_1,\tilde{J}^{c_2}_n=-x_2,\cdots,\tilde{J}^{c_{r}}_n=-x_{r}\right)}
		= e^{n\sum_{i=1}^{r}x_i\log\frac{\gamma^{c_i}}{\gamma^{c_i-}}}.
	\end{split}
\end{equation}
Note that \eqref{weak} is much weaker than \eqref{strong}. In what follows, we term \eqref{strong} the \emph{strong form} and term \eqref{weak} the \emph{weak form} of the transient fluctuation theorem.

Other types of fluctuation theorems for absolute and net LE currents can be easily derived from the transient fluctuation theorem and are summarized as follows. Here we only focus on the strong form of various fluctuation theorems; the weak form can be obtained similarly. Let $g_n(\lambda) = \langle e^{n\sum_{c\in\mathcal{C}}\lambda_iJ^{c_i}_n}\rangle$ and $\tilde{g}_n(\lambda) = \langle e^{n\sum_{i=1}^{r}\lambda_i\tilde{J}^{c_i}_n}\rangle$ be the moment generating functions of $(J^{c_i}_n)_{c_i\in\mathcal{C}}$ and $(\tilde{J}^{c_i}_n)_{1\le i\le r}$, respectively. Moreover, let $I_{J}(x)$ and $I_{\tilde{J}}(x)$ be the rate functions of $(J^{c_i}_n)_{c_i\in\mathcal{C}}$ and $(\tilde{J}^{c_i}_n)_{1\le i\le r}$, respectively.

1) Kurchan-Lebowitz-Spohn-type fluctuation theorem: if cycles $c_1$ and $c_2$ are similar, then
\begin{equation*}
	g_n(\lambda_1,\lambda_2,\cdots) = g_n\left(\lambda_2-\log\frac{\gamma^{c_1}}{\gamma^{c_2}},\lambda_1+\log\frac{\gamma^{c_1}}{\gamma^{c_2}},\cdots\right).
\end{equation*}
\begin{equation*}
	\tilde{g}_n(\lambda_1,\cdots)=\tilde{g}_n\left(-\left(\lambda_1+\log\frac{\gamma^{c_1}}{\gamma^{c_1-}}\right),\cdots\right).
\end{equation*}

2) Integral fluctuation theorem: for any subset $\{c_1,c_2,\cdots,c_t\}\subset\{c_1,c_2,\cdots,c_r\}$, we have
\begin{equation}\label{integral}
	\left\langle e^{-n\sum_{i=1}^t\tilde{J}^{c_i}_n\log\frac{\gamma^{c_i}}{\gamma^{c_i-}}}\right\rangle=1.
\end{equation}

3) Gallavotti-Cohen-type fluctuation theorem: if cycles $c_1$ and $c_2$ are similar, then
\begin{equation*}
	I_J(x_1,x_2,\cdots)=I_J(x_2,x_1,\cdots)-\left(\log\frac{\gamma^{c_1}}{\gamma^{c_2}}\right)(x_1-x_2).
\end{equation*}
\begin{equation*}
	I_{\tilde{J}}(x_1,\cdots) = I_{\tilde{J}}(-x_1,\cdots)-\left(\log\frac{\gamma^{c_1}}{\gamma^{c_1-}}\right)x_1.
\end{equation*}

The fluctuation theorems for net LE currents have important physical implications. To see this, recall that the total entropy production of a Markovian system along a single trajectory $\{\xi_0,\xi_1,\cdots,\xi_n\}$ is given by \cite{seifert2005entropy}
	\begin{equation*}
		S^{tot}_n = \log\frac{\mu_0(\xi_0)p_{\xi_0\xi_1}p_{\xi_1\xi_2}\cdots p_{\xi_{n-1}\xi_n}}
		{\mu_n(\xi_n)p_{\xi_n\xi_{n-1}}p_{\xi_{n-1}\xi_{n-2}}\cdots p_{\xi_1\xi_0}}
		= \log\frac{\mu_0(\xi_0)}{\mu_n(\xi_n)}+\sum_{k=0}^{n-1}\frac{p_{\xi_k\xi_{k+1}}}{p_{\xi_{k+1}\xi_k}},
	\end{equation*}
	where $\mu_0 = (\mu_0(i))_{i\in S}$ is the distribution of $\xi_0$ and $\mu_n = (\mu_n(i))_{i\in S}$ is the distribution of $\xi_n$. Under the periodic boundary condition, it is clear that $\mu_{0}(\xi_0)=\mu_{n}(\xi_n)$. Moreover, we have
	\begin{equation*}
		p_{\xi_0\xi_1}p_{\xi_1\xi_2}\cdots p_{\xi_{n-1}\xi_n}=\prod_{c\in\mathcal{C}}(\gamma^c)^{N^c_n}, \qquad p_{\xi_n\xi_{n-1}}p_{\xi_{n-1}\xi_{n-2}}\cdots p_{\xi_1\xi_0}=\prod_{c\in\mathcal{C}}(\gamma^c)^{N^{c-}_n}.
	\end{equation*}
	Combining the above two equations, we obtain
	\begin{equation}\label{entropy production}
		S^{tot}_n = n\sum_{c\in\mathcal{C}}\tilde{J}^c_n\log\gamma^c
		= \frac{n}{2}\sum_{c\in\mathcal{C}}\tilde{J}^c_n\log\frac{\gamma^c}{\gamma^{c-}}
		= n\sum_{i=1}^r\tilde{J}^{c_i}_n\log\frac{\gamma^{c_i}}{\gamma^{c_i-}},
	\end{equation}
	where we have used the fact that $\tilde{J}^{c-}_n = -\tilde{J}^c_n$ in the second identity. This shows that the total entropy production can be decomposed as the weighted sum of net LE currents with the weights being all cycle affinities, and the quantity $n\tilde{J}^c_n\log(\gamma^c/\gamma^c)$ can be understood as the entropy product along cycle $c$. It is well known as the total entropy production of any Markovian system satisfies the integral fluctuation theorem $\langle e^{-S^{tot}_n} \rangle= 1$ \cite{seifert2005entropy}, which implies the classical second law of thermodynamics $\langle S^{tot}_n\rangle\geq 0$. Our results indicate that the integral fluctuation theorem not only holds for the total entropy production, but also holds for the entropy production along any finite number of cycles $c_1,c_2,\cdots,c_t$ (see \eqref{integral}). In particular, for any cycle $c$, we have
	\begin{equation*}
		\left\langle e^{-n\tilde{J}^c_n\log\frac{\gamma^c}{\gamma^{c-}}}\right\rangle=1.
	\end{equation*}
	This much stronger than the classical result for the total entropy production. Moreover, applying Jensen's inequality to the integral fluctuation theorem \eqref{integral}, we find
	\begin{equation}\label{secondlaw}
		\left\langle\sum_{i=1}^t\tilde{J}^{c_i}_n\log\frac{\gamma^{c_i}}{\gamma^{c_i-}}\right\rangle\geq 0.
	\end{equation}
	where $c_1,c_2,\cdots,c_t$ are any finite number of cycles. In particular, for any cycle $c$, we have
	\begin{equation*}
		\left\langle\tilde{J}^c_n\log\frac{\gamma^c}{\gamma^{c-}}\right\rangle\geq 0.
	\end{equation*}
	This provides a much refined version of the second law of thermodynamics, which shows that the entropy production along any finite number of cycles has a nonnegative mean. This reveals the hidden refined structure behind the underlying system.

\subsection{Fluctuation theorems for ST currents of general Markov chains}\label{sec:FT ST}
\begin{figure}[h]
	\centering\includegraphics[width=0.8\textwidth]{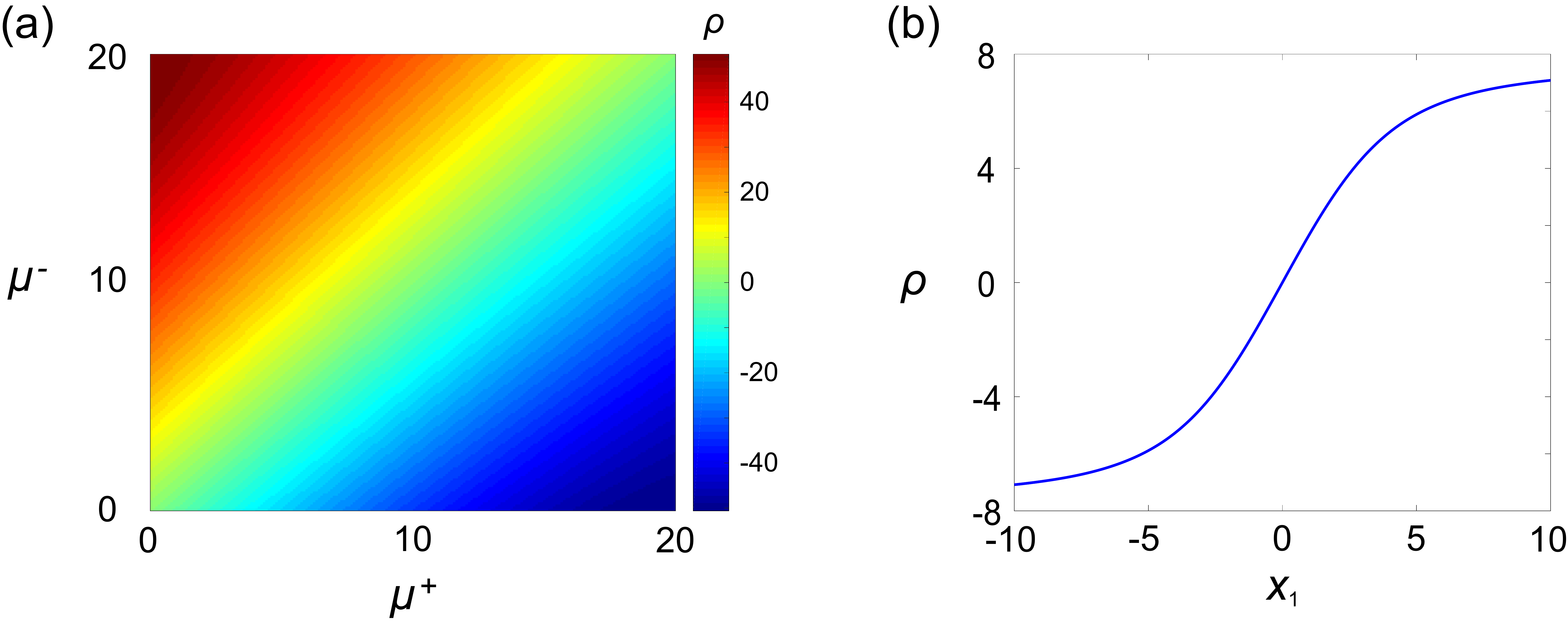}
	\caption{\textbf{Some fluctuation theorems may be broken for absolute and net ST currents.} (\textbf{a}) Heat plot of $\rho=I_Q(\mu^+,\mu^-,\cdots)-I_Q(\mu^-,\mu^+,\cdots)+\log(\gamma^+/\gamma^-)(\mu^+-\mu^-)$ as a function of $\mu^+$ and $\mu^-$ for a three-state system. The fact that $\rho\not\equiv 0$ shows that the Gallavotti-Cohen-type fluctuation theorem is broken for absolute ST currents. The parameters are chosen as $\mu^1=15$, $\mu^2=20$, $\mu^3=3$, $\mu^{12}=21$, $\mu^{23}=37$, $p_{11}=0.28$, $p_{12}=0.22$, $p_{13}=0.5$, $p_{21}=0.1$, $p_{22}=0.6$, $p_{23}=0.3$, $p_{31}=0.3$, $p_{32}=0.3$, $p_{33}=0.4$. (\textbf{b}) Change of $\rho=I_{\tilde{Q}}(x_1,x_2,x_3)- I_{\tilde{Q}}(-x_1,x_2,x_3)
		+\log(\gamma^{c_1}/\gamma^{c_1-})x_1$ as a function of $x_1$ for a four-state system. The fact that $\rho\not\equiv 0$ shows that the strong form of the Gallavotti-Cohen-type fluctuation theorem is broken for net ST currents. The parameters are chosen as $x_2=2$, $x_3=3$, $p_{11}=0.1$, $p_{12}=0.2$, $p_{13}=0.3$, $p_{14}=0.4$, $p_{21}=0.5$, $p_{22}=0.15$, $p_{23}=0.15$, $p_{24}=0.2$, $p_{31}=0.1$, $p_{32}=0.4$, $p_{33}=0.25$, $p_{34}=0.25$, $p_{41}=0.2$, $p_{42}=0.2$, $p_{43}=0.3$, $p_{44}=0.3$.}\label{figure:ratefunction}
\end{figure}

We have seen that both absolute and net LE currents satisfy various fluctuation theorems. A natural question is whether similar relations also hold for absolute and net ST currents. In fact, (absolute) ST currents do not satisfy any form of fluctuation theorems, even for monocyclic systems. To see this, consider a fully connected three-state system and let $T = 1\to 2\to 3$ be the spanning tree. Then the fundamental set is given by
\begin{equation*}
	\mathcal{L} = \{(1),(2),(3),(1,2),(2,3),(1,2,3),(1,3,2)\}.
\end{equation*}
It then follows from \eqref{formula:I_Q} that the rate function for empirical ST currents is given by
\begin{equation*}
	I_Q(\mu) = \sum_{\langle i,j\rangle\in E}R^{\mu}(i,j)\log\frac{R^{\mu}(i,j)}{R^{\mu}(i)p_{ij}},
\end{equation*}
where $R^{\mu}(i,j)=\sum_{c_l\in\mathcal{L}}\mu^{c_l}H^{c_l}(i,j)$ and $R^{\mu}(i)=\sum_{j\in S}R^{\mu}(i,j)$. For simplicity of notation, let $\mu^+ = \mu^{(1,2,3)}$ and let $\mu^- = \mu^{(1,3,2)}$. In Fig. \ref{figure:ratefunction}(a), we illustrate the difference between $I_Q(\mu^+,\mu^-,\cdots)$ and $I_Q(\mu^-,\mu^+,\cdots)-\log(\gamma^+/\gamma^-)(\mu^+-\mu^-)$ as a function of $\mu^+$ and $\mu^-$ under a set of appropriately chosen parameters. It is clear that the difference is nonzero and thus we have
\begin{equation}\label{ratefunction3}
	I_Q(\mu^+,\mu^-,\cdots)
	\neq I_Q(\mu^-,\mu^+,\cdots)-\left(\log\frac{\gamma^+}{\gamma^-}\right)(\mu^+-\mu^-),
\end{equation}
which means that the Gallavotti-Cohen-type fluctuation theorem is broken. Other types of fluctuation theorems must also be broken since the Gallavotti-Cohen-type fluctuation theorem is the weakest among all fluctuation theorems.

While various fluctuation theorems fail for ST currents, they may hold for net ST currents \cite{andrieux2007fluctuation}. To see this, note that for a monocyclic system, we only need to consider the empirical net current $\tilde{Q}^+_n$ of cycle $C^+$. Suppose that the spanning tree is chosen as $T = 1\to 2\to\cdots \to N$. With the periodic boundary condition, it follows from \eqref{conversion} that $Q^+_n = J^+_n+J^{(N,1)}_n$ and $Q^-_n = J^-_n+J^{(N,1)}_n$. These two equations imply that $\tilde{Q}^+_n = \tilde{J}^+_n$, and thus the fluctuation theorems for net ST currents naturally follow from those for net LE currents. Without the periodic boundary condition, the Gallavotti-Cohen-type fluctuation theorem still holds since it reflects the long-time behavior of the system and assuming the periodic boundary condition or not will not influence the large deviation rate function, while the other three types of fluctuation theorems are all broken. It has been shown in \cite{polettini2014transient} that all the four types of fluctuation theorems are satisfied for a modified version of net ST currents.

The above results can be extended to a general system. Let $c_{l_1},c_{l_2},\cdots,c_{l_s}$ be all cycles with three or more states in the fundamental set so that any two of them are not reversed cycles of each other (the empirical net ST currents for one-state and two-state cycles vanish and do not need to be considered). In \cite{andrieux2007fluctuation}, the authors have proved the following weak form of the Gallavotti-Cohen-type fluctuation theorem for net ST currents:
\begin{equation}\label{ST net LDP flu}
	I_{\tilde{Q}}(x_1,x_2,\cdots,x_s)
	= I_{\tilde{Q}}(-x_1,-x_2,\cdots,-x_s)
	-\sum_{i=1}^sx_i\log\frac{\gamma^{c_{l_i}}}{\gamma^{c_{l_i-}}}.
\end{equation}
This shows that the joint distribution of empirical net ST currents satisfies a symmetric relation when all $x_i$ are replaced by $-x_i$. In fact, the above equality can be obtained directly from the fluctuation theorems for net LE currents. For any cycle $c_l\in\mathcal{L}$ with three or more states, under the periodic boundary condition, it follows from \eqref{conversion} that
\begin{align}\label{tilde Q}
	\tilde{Q}^{c_l}_n &= \sum_{c\ni l}J^c_n-\sum_{c\ni l-}J^c_n
	= \sum_{c\ni l}J^c_n-\sum_{c\ni l}J^{c-}_n = \sum_{c\ni l}\tilde{J}^c_n.
\end{align}
This indicates that empirical net ST currents can be decomposed as the sum of empirical net LE currents. It then follows from \eqref{weak} that (see Appendix~\ref{appendix?B} for the proof)
\begin{align}\label{transient fluctuation theorem}
	\frac{\mathbb{P}\left(\tilde{Q}^{c_{l_1}}_n=x_1,\cdots,\tilde{Q}^{c_{l_s}}_n=x_s\right)}
	{\mathbb{P}\left(\tilde{Q}^{c_{l_1}}_n=-x_1,\cdots,\tilde{Q}^{c_{l_s}}_n=-x_s\right)}
	= e^{n\sum_{i=1}^sx_i\log\frac{\gamma^{c_{l_i}}}{\gamma^{c_{l_i}-}}}.
\end{align}
This shows that net ST currents satisfy the weak form of the transient fluctuation theorem under the periodic boundary condition. The weak form \eqref{ST net LDP flu} of the Gallavotti-Cohen-type fluctuation theorem holds generally since assuming the periodic boundary condition or not will not influence the large deviation rate function.

In contrast to net LE currents, net ST currents do not satisfy the strong form of fluctuation theorems; this fact has been found in previous papers \cite{mehl2012role, polettini2017effective, uhl2018fluctuations, kahlen2018hidden}. To give a counterexample, we consider a fully connected four-state system illustrated in Fig. \ref{figure:transitiongraph}(b). Suppose that the spanning tree is chosen as $T=1\to2\to3\to4$. In this case, we only need to consider the net ST currents of the three cycles $c_1 = (1,2,3)$, $c_2 = (2,3,4)$, and $c_3 = (1,2,3,4)$, since other cycles in the fundamental set are either their reversed cycles or cycles with one or two states. Recall that the rate function of empirical net ST currents $(\tilde{Q}^{c_1}_n,\tilde{Q}^{c_2}_n,\tilde{Q}^{c_3}_n)$ is given by
\begin{equation}\label{ratefunction5}
	I_{\tilde{Q}}(x)=\inf_{\{\mu\in\mathcal{M}:\mu^{c_i}-\mu^{c_i-}= x_i,\;\forall 1\le i\le 3\}}I_Q(\mu).
\end{equation}
In Fig. \ref{figure:ratefunction}(b), we illustrate the difference between $I_{\tilde{Q}}(x_1,x_2,x_3)$ and $I_{\tilde{Q}}(-x_1,x_2,x_3)-\log(\gamma^{c_1}/\gamma^{c_1-})\tilde{\mu}^{c_1}$ as a function of $x_1$ under a set of appropriately chosen parameters. It is clear that the difference is nonzero and thus
\begin{equation}\label{ratefunction4}
	I_{\tilde{Q}}(x_1,x_2,x_3)\neq I_{\tilde{Q}}(-x_1,x_2,x_3)
	-\left(\log\frac{\gamma^{c_1}}{\gamma^{c_1-}}\right)\tilde{\mu}^{c_1}.
\end{equation}
Hence the strong form of the Gallavotti-Cohen-type fluctuation theorem fails for net ST currents.

We next discuss the connection between ST currents and entropy production. Similarly to  \eqref{entropy production}, under the periodic boundary condition, the total entropy product along a single trajectory can be also decomposed as the weighted sum of net ST currents \cite{schnakenberg1976network}, i.e.
	\begin{equation}\label{entropyST}
		S^{tot}_n = n\sum_{i=1}^s\tilde{Q}^{c_{l_i}}_n\log\frac{\gamma^{c_{l_i}}}{\gamma^{c_{l_i-}}}.
	\end{equation}
	Hence within the spanning tree framework, the quantity $n\tilde{Q}^{c_l}_n\log(\gamma^{c_l}/\gamma^{c_l-})$ can be understood as the entropy production \emph{along fundamental cycle} $c_l$. Note that this is totally different from the quantity $\tilde{J}^{c_l}_n\log(\gamma^{c_l}/\gamma^{c_l-})$ investigated in Section \ref{sec:FT ST}. We have seen that within the loop-earased framework, the entropy production along any finite number of cycles satisfies both the strong form of integral fluctuation theorem \eqref{integral} and the refined version of the second law of thermodynamics \eqref{secondlaw}. Since the strong form of fluctuation theorems fails for net ST currents, the entropy production \emph{along any fundamental cycle} does not satisfy the refined version of the second law of thermodynamics. In other words, it may occur that
	\begin{equation*}
		\left\langle\tilde{Q}^{c_l}_n\log\frac{\gamma^{c_l}}{\gamma^{c_l-}}\right\rangle < 0,
	\end{equation*}
	for some fundamental cycle $c_l$.
	
	The reason why the strong form of fluctuation theorems and the refined version of the second law of thermodynamics are broken for net ST currents can be explained as follows. From \eqref{tilde Q}, it is clear that the net ST current $\tilde{Q}^{c_l}_n$ of fundamental cycle $c_l$ can be decomposed as the sum of the net LE currents $\tilde{J}^c_n$ of all cycles $c$ that traverse chord $l$, i.e. $\tilde{Q}^{c_l}_n = \sum_{c\ni l}\tilde{J}^c_n$. Note that these cycles $c$ that traverse chord $l$ have different affinities, which may not be equal to the affinity of fundamental cycle $c_l$. Hence even if $\langle\tilde{J}^c_n\log(\gamma^c/\gamma^{c-})\rangle\geq 0$ for all cycle $c$, we cannot conclude that $\langle\tilde{Q}^{c_l}_n\log(\gamma^{c_l}/\gamma^{c_l-})\rangle\geq 0$. The weak form of fluctuation theorems holds for net ST currents since it is essentially the fluctuation theorems for the total entropy production (see \eqref{entropyST}).
	
	In summary, we have seen that LE currents have much better properties than ST currents; the former satisfies a much refined version of the second law of thermodynamics while the latter does not. This demonstrates the advantage of LE currents in dealing with complex thermodynamic systems far from equilibrium (for simple monocyclic systems, the net LE and ST currents are the same).

\section{Conclusions and discussion}
In this paper, we make a comparative study between the large deviations and fluctuation theorems for empirical cycle currents of a Markov chain defined in the LE and ST manners. LE currents are defined for all cycles in the cycle space, while ST currents are only defined for cycles in the fundamental set generated by the chords of an arbitrarily chosen spanning tree. The fundamental set may be much smaller than the cycle space for a general system. However, for a system with a cyclic topology, there is at most one cycle that is contained in the cycle space but is missing in the fundamental set. LE currents provide a more complete and detailed description for the cycle dynamics than ST currents. Under the periodic boundary condition, the ST current of any cycle can be represented by the weighted sum of LE currents.

Furthermore, we establish the large deviation principle and provide the explicit expression of the associated rate function for empirical LE currents of a monocyclic Markov chain. The proof is based on deriving the joint distribution of empirical LE currents of all cycles in closed form. When computing the joint distribution, we propose the method of three-step cycle insertion: (i) the first step is to insert all cycles that pass through the initial state into the trajectory, (ii) the second step is to insert all two-state cycles that do not contain the initial state into the trajectory, (iii) and the third step is to insert all one-state cycles that do not contain the initial state into the trajectory. In addition, the rate function is proved to be independent of the initial distribution of the system. The analytical expression of the rate function is complicated for a general monocyclic system. However, it can be greatly simplified for a three-state system and for a monocyclic system with a certain transition between adjacent states being forbidden. Following the method proposed in \cite{bertini2015flows} which only focused on empirical net ST currents, we also give the exact rate function for empirical (absolute) ST currents of a general system. The relationship between the rate functions of empirical LE and ST currents is clarified.

The analytical results are then applied to investigate the fluctuations in the product rate for a three-step reversible enzyme reaction, which can be modelled as a three-state monocyclic system. A single enzyme molecule can convert stochastically among three conformal states: the free enzyme $E$, the enzyme-substrate complex $ES$, and the enzyme-product complex $EP$. The product rate of the enzyme reaction is exactly the empirical net LE current of the monocyclic system. Using the exact joint distribution for LE currents, we find that rapid equilibrium between the enzyme states $E$ and $ES$ and rapid equilibrium between the enzyme states $ES$ and $EP$ both result in larger fluctuations in the product rate. Moreover, compared with the analytical solution, we show that the large deviations for LE currents provide a much more efficient computational method of the joint distribution, and thus enables a fast exploration of large swaths of parameter space.

Finally, we examine various types of fluctuation theorems satisfied by empirical LE and ST currents and clarify their ranges of applicability. We first show that the empirical absolute and net LE currents satisfy all types of fluctuation theorems and symmetric relations. In particular, we introduce the concept of similar cycles and obtain the strong form of the transient fluctuation theorem: (i) the joint distribution of empirical LE currents satisfies a symmetric relation when the currents of any pair of similar cycles are exchanged; (ii) the joint distribution of empirical net LE currents satisfies a symmetry relation when the net current of any cycle is replaced by its opposite number. Since empirical ST currents can be represented by the weight sums of empirical LE currents under the periodic boundary condition, we further show that empirical ST currents do not satisfy any form of fluctuation theorems, while empirical net ST currents only satisfy the weak form of the transient fluctuation theorem under the periodic boundary condition: the joint distribution of empirical net ST currents satisfies a symmetry relation when the net currents of all cycles in the fundamental set are replaced by their opposite numbers. As a corollary of the integral fluctuation theorem, we show that LE currents satisfy a refined version of the second law of thermodynamics: the entropy production along any finite number of cycles has a nonnegative mean, while it is broken for ST currents.

In the present paper, some results are only obtained for a monocyclic Markov chain. We anticipate that these results can be generalized to more general Markovian systems and even to semi-Markovian or non-Markovian systems. In addition, here we only make a comparison between LE and ST currents. The relationship between these two types of cycle currents and those defined in the sequence matching manner \cite{pierpaolo2019exact, john2020reversal, patrick2021cycle} is not clear. These are under current investigation.

\section*{Acknowledgements}
We are grateful to Professor H. Qian for stimulating discussions. C.\ J.\ acknowledges support from National Natural Science Foundation of China with grant No. U1930402 and grant No. 12131005.

\begin{appendices}

	\section{Simplified expression of rate function $I_J$ in two special cases}\label{appendix:threestate}
	We have seen that the rate function $I_J$ for empirical LE currents of a monocyclic system can be simplified to a large extent in two special cases: (i) the case where the system has only three states and (ii) the case where the transition from state 1 to state $N$ is forbidden (see Fig. \ref{figure:transitiongraph}(d) for an illustration). Next we will give the proof.
	
	We first prove that for a three-state system, the rate function is given by \eqref{ratefuntion 3state}. When $N = 3$, it is easy to see that the solution $X=(x^2,y^2)$ of \eqref{equation} is given by
	\begin{equation*}
		x^{2}=\frac{\nu^{23}\left(\nu^{12}+\nu^+\right)}{\nu^{12}+\nu^{13}+\nu^++\nu^-},\quad y^{2}=\frac{\nu^{23}\left(\nu^{13}+\nu^-\right)}{\nu^{12}+\nu^{13}+\nu^++\nu^-}.
	\end{equation*}
	Note that the solution $X=(x^2,y^2)$ minimizes the function $F_{\nu}$. Then we have
	\begin{equation*}\label{Fnu2}
		I_2(\nu)=F_{\nu}(X)=\nu^{23}\log\frac{\nu^{23}}{\tilde{\nu}}+\left(\nu^{12}+\nu^{13}+\nu^++\nu^-\right)\log\frac{\tilde{\nu}-\nu^{23}}{\tilde{\nu}}.
	\end{equation*}
	Straightforward calculations show that
	\begin{equation}\label{I1I2I3}
		I_1(\nu)+I_2(\nu)+I_3(\nu) = \sum_{i\in S} \left[\nu^{i}\log \frac{\nu^{i}}{\nu_i} + (\nu_i - \nu^i)\log \frac{\nu_i - \nu^i}{\nu_i}
		\right]
		+ \sum_{c \in \mathcal{C}, |c|\neq 1} \nu^{c} \log \frac{\nu^{c}}{\tilde{\nu}}.
	\end{equation}
	Recall the following expression of the LE currents \cite[Theorem.1.3.3]{jiang2004mathematical}:
	\begin{equation}\label{J+J-Jii+1}
		J^+=\gamma^+\frac{1}{C},\quad J^-=\gamma^-\frac{1}{C},\quad J^{i,i+1}=\gamma^{i,i+1}\frac{1-p_{i-1,i-1}}{C}, \quad 1\le i\le 3,
	\end{equation}
	where $C=\sum_{i\in S}[(1-p_{i-1,i-1})(1-p_{i+1,i+1})-p_{i-1,i+1}p_{i+1,i-1}]$. It then follows from \eqref{decomposition} that
	\begin{equation}\label{pij}
		p_{ij}=\frac{\sum_{c \ni \langle i,j\rangle}J^c}{\sum_{c\ni i}J^c}.
	\end{equation}
	Combining \eqref{J+J-Jii+1} and \eqref{pij}, we have
	\begin{equation}\label{nucgamma2}
		\begin{split}
			&\;\sum_{i \in S}\left[\nu^i\log\frac{J^i}{J_i}+\left(\nu_i-\nu^i\right)\log\frac{J_i-J^i}{J_i}\right]+ \sum_{c \in \mathcal{C}, |c|\neq 1} \nu^{c} \log \left(\frac{J^c}{\tilde{J}}\right)\\
			=&\;\sum_{i \in S}\left[\nu^i\log\frac{J^i}{J_i}+\nu^{i,i+1}\log\left(\left(1-\frac{J^i}{J_i}\right)\left(1-\frac{J^{i+1}}{J_{i+1}}\right)\frac{J^{i,i+1}}{\tilde{J}}\right)\right]\\
			&\;+\nu^+\log\left(\frac{J^+}{\tilde{J}}\prod_{i\in\mathcal{C}}\left(1-\frac{J^i}{J_i}\right)\right)+\nu^-\log\left(\frac{J^-}{\tilde{J}}\prod_{i\in\mathcal{C}}\left(1-\frac{J^i}{J_i}\right)\right)\\
			=&\;\sum_{c \in \mathcal{C}}\nu^c \log\gamma^c=-I_4(\nu).
		\end{split}
	\end{equation}
	Combining \eqref{I1I2I3} and \eqref{nucgamma2} gives the desired result.
	
	We next prove that for a monocyclic system, if the transition from state 1 to state $N$ is forbidden (see Fig. \ref{figure:transitiongraph}(d) for an illustration), then the rate function is given by \eqref{forbidden}. Since $p_{N1}=0$, the two cycles $(1,N)$ and $(1,N,\cdots,2)$ cannot be formed. Hence we can take $\nu^{N1}=\nu^{-}=0$ in \eqref{equation} and it is easy to see that $x^{i}=\nu^{i,i+1}$, $y^{i}=0$ is a solution of \eqref{equation}. Then we have
	\begin{equation*}
		I_2(\nu)=F_{\nu}(x^i,y^i)=\sum_{i=2}^{N-1}\left[-\lambda_{i}\nu^{i,i+1}+\nu^+\log\frac{\nu^{i-1,i}+\nu^+}{\nu^{i-1,i}+\nu^{i,i+1}+\nu^+}\right]+\nu^{12}\log\frac{\nu^{12}+\nu^+}{\nu^{12}+\nu^{23}+\nu^+},
	\end{equation*}
	where
	\begin{equation*}
		\lambda_i=-\log\left(\frac{\nu^{i,i+1}}{\nu^{i-1,i}+\nu^{i,i+1}+\nu^+}\,\frac{\nu^{i,i+1}+\nu^+}{\nu^{i,i+1}+\nu^{i+1,i+2}+\nu^+}\right).
	\end{equation*}
	By the definition of $\nu_i$, we have
	\begin{gather*}
		\nu_1=\nu^{1}+\nu^{12}+\nu^+,\\
		\nu_i=\nu^i+\nu^{i-1,i}+\nu^{i,i+1}+\nu^+,\;\;\;2\le i\le N-1,\\
		\nu_N=\nu^N+\nu^{N-1,N}+\nu^+.
	\end{gather*}
	Straightforward calculations show that
	\begin{gather}
		I_1(\nu) = \nu^{12}\log\frac{\nu^{12}}{\nu_1-\nu^{12}}+\nu^+\log\frac{\nu^+}{\nu_1-\nu^1},\nonumber\\
		I_2(\nu) =\sum_{i=2}^{N}\left[\nu^{i,i+1}\log\frac{\nu^{i,i+1}}{\nu_i-\nu^i}+\nu^+\log\frac{\nu^{i-1,i}+\nu^+}
		{\nu_i-\nu^i}\right]+\sum_{i=1}^{N}\nu^{i,i+1}\log\frac{\nu^{i,i+1}+\nu^+}{\nu_{i+1}-\nu^{i+1}},\label{threeeqns}\\
		I_3(\nu) = \sum_{i\in S}\left[\nu^i\log\frac{\nu^i}{\nu_i}+\nu^+\log\frac{\nu_i-\nu^i}{\nu_i}\right]+\sum_{i\in S}\nu^{i,i+1}\left(\log\frac{\nu_i-\nu^i}{\nu_i}
		+\log\frac{\nu_{i+1}-\nu^{i+1}}{\nu_{i+1}}\right).\nonumber
	\end{gather}
	It then follows from \eqref{pij} that
	\begin{equation}\label{nucgammac}
		\begin{split}
			&\;\sum_{i \in S}\left[\nu^i\log\frac{J^i}{J_i}+\nu^{i,i+1}\log\frac{J^{i,i+1}}{J_i}+(\nu^{i-1,i}+\nu^+)\log\frac{J^{i-1,i}+J^+}{J_i}\right]\\
			=&\;\sum_{i \in S}\left[\nu^i\log\frac{J^i}{J_i}+\nu^{i,i+1}\log\frac{J^{i,i+1}(J^{i,i+1}+J^+)}{J_{i+1}J_i}\right]+\nu^+\log\frac{\prod_{i=1}^N\left(J^{i,i+1}+J^+\right)}{\prod_{i=1}^N J_i}\\
			=&\;\sum_{c \in \mathcal{C}}\nu^c \log\gamma^c=-I_4(\nu).
		\end{split}
	\end{equation}
	Combining \eqref{threeeqns} and \eqref{nucgammac} gives the desired result.
	
	\section{Proof of the transient fluctuation theorem for net ST currents}\label{appendix?B}
	Here we will prove \eqref{transient fluctuation theorem} under the periodic boundary condition. It follows from \eqref{tilde Q} that for any cycle $c_l\in\mathcal{L}$ with three or more states, we have
	\begin{equation*}
		\tilde{Q}^{c_l}_n=\sum_{i=1}^r\tilde{J}^{c_i}_n[1_{\{l\in c_i\}}-1_{\{l\in c_i-\}}],
	\end{equation*}
	where $1_A$ is the indicator function which takes the value of $1$ when $A$ holds and takes the value of $0$ when $A$ does not hold. Then we obtain
	\begin{equation*}
		{\small\begin{split}
				&\;\mathbb{P}\left(\tilde{Q}^{c_{l_1}}_n=x_1,\cdots,\tilde{Q}^{c_{l_s}}_n=x_s\right)\\
				=&\;\mathbb{P}\left(\sum_{i=1}^r\tilde{J}^{c_i}_n\left[1_{\{l_1\in c_i\}}-1_{\{l_1\in c_i-\}}\right]=x_1,\cdots,\sum_{i=1}^r\tilde{J}^{c_i}_n\left[1_{\{l_s\in c_i\}}-1_{\{l_s\in c_i-\}}\right]=x_s\right)\\
				=&\;\sum_{\sum_{i=1}^ry_i[1_{\{l_m\in c_i\}}-1_{\{l_m\in c_i-\}}]=x_m,1\le m\le s}\mathbb{P}\left(\tilde{J}^{c_1}_n=y_1,\cdots,\tilde{J}^{c_r}_n=y_r\right)\\
				=&\;\sum_{\sum_{i=1}^ry_i[1_{\{l_m\in c_i\}}-1_{\{l_m\in c_i-\}}]=x_m,1\le m\le s}\mathbb{P}\left(\tilde{J}^{c_1}_n=-y_1,\cdots,\tilde{J}^{c_r}_n=-y_r\right)e^{n\sum_{i=1}^ry_i\log\frac{\gamma^{c_i}}{\gamma^{c_i-}}}\\
				=&\;\sum_{\sum_{i=1}^ry_i[1_{\{l_m\in c_i\}}-1_{\{l_m\in c_i-\}}]=x_m,1\le m\le s}\mathbb{P}\left(\tilde{J}^{c_1}_n=-y_1,\cdots,\tilde{J}^{c_r}_n=-y_r\right)e^{n\sum_{i=1}^sx_i\log\frac{\gamma^{c_{l_i}}}{\gamma^{c_{l_i}-}}}\\
				=&\;\sum_{\sum_{i=1}^ry_i[1_{\{l_m\in c_i\}}-1_{\{l_m\in c_i-\}}]=-x_m,1\le m\le s}\mathbb{P}\left(\tilde{J}^{c_1}_n=y_1,\cdots,\tilde{J}^{c_r}_n=y_r\right)e^{n\sum_{i=1}^sx_i\log\frac{\gamma^{c_{l_i}}}{\gamma^{c_{l_i}-}}}\\
				=&\;\mathbb{P}\left(\sum_{i=1}^r\tilde{J}^{c_i}_n\left[1_{\{l_1\in c_i\}}-1_{\{l_1\in c_i-\}}\right]=-x_1,\cdots,\sum_{i=1}^r\tilde{J}^{c_i}_n\left[1_{\{l_s\in c_i\}}-1_{\{l_s\in c_i-\}}\right]=-x_s\right)e^{n\sum_{i=1}^sx_i\log\frac{\gamma^{c_{l_i}}}{\gamma^{c_{l_i}-}}}\\
				=&\;\mathbb{P}\left(\tilde{Q}^{c_{l_1}}_n=x_1,\cdots,\tilde{Q}^{c_{l_s}}_n=x_s\right)e^{n\sum_{i=1}^sx_i\log\frac{\gamma^{c_{l_i}}}{\gamma^{c_{l_i}-}}},
		\end{split}}
	\end{equation*}
	where we use the fact that under the constraint of $\sum_{i=1}^ry_i[1_{\{l_m\in c_i\}}-1_{\{l_m\in c_i-\}}]=x_m,\;\forall\,1\le m\le s$, we have
	\begin{equation}\label{1}
		\sum_{i=1}^ry_i\log\frac{\gamma^{c_i}}{\gamma^{c_i-}}=\sum_{j=1}^sx_j\log\frac{\gamma^{c_{l_j}}}{\gamma^{c_{l_j}-}}.
	\end{equation}
	
	This identity is highly nontrivial and we next prove it. For any cycle $c$, let $L^c$ be a function on $E$ defined by
	\begin{equation*}\label{cycle function}
		L^c(i,j)
		=\left\{\begin{aligned}
			1, &   && \text{if } \langle i,j\rangle \in c,\\
			0, &   && \text{otherwise}.\\
		\end{aligned}\right.
	\end{equation*}
	By the definition of the function $H^{c_l}$ in \eqref{cycle function2}, it can be proved that \cite{kalpazidou2007cycle}
	\begin{equation*}
		L^c = \sum_{l\notin T}L^c(l)H^{c_l}.
	\end{equation*}
	Let $w$ be a function on $E$ defined by
	\begin{equation*}
		w(i,j)=\log\frac{p_{ij}}{p_{ji}}.
	\end{equation*}
	For any cycle $c=(i_1,i_2,\cdots,i_t)$, we have
	\begin{equation*}
		\log\frac{\gamma^{c}}{\gamma^{c-}} = \sum_{k=1}^t\log\frac{p_{i_k,i_{k+1}}}{p_{i_{k+1},i_k}}
		= \langle w,L^c\rangle,
	\end{equation*}
	where $i_{t+1}=i_1$ and $\langle w,L^c\rangle = \sum_{\langle i,j\rangle\in E}w(i,j)L^c(i,j)$ is the inner product. Moreover, for any $c_l\in\mathcal{L}$, it is not difficult to prove that
	\begin{equation*}
		\log\frac{\gamma^{c_l}}{\gamma^{c_l-}}=\langle w,H^{c_l}\rangle.
	\end{equation*}
	Note that $\log(\gamma^{c}/\gamma^{c-})=0$ for all one-state or two-state cycles. Then for any cycle $c$, we have
	\begin{equation*}
		\sum_{j=1}^s\left[L^c(l_j)-L^{c-}(l_j)\right]\log\frac{\gamma^{c_{l_j}}}{\gamma^{c_{l_j}-}}
		=\sum_{l\notin T}L^c(l)\log\frac{\gamma^{c_{l}}}{\gamma^{c_{l}-}}.
	\end{equation*}
	Thus we finally obtain
	\begin{align*}
		\sum_{j=1}^sx_j\log\frac{\gamma^{c_{l_j}}}{\gamma^{c_{l_j}-}}&=\sum_{j=1}^s\sum_{i=1}^ry_i\left[L^{c_i}(l_j)-L^{c_i-}(l_j)\right]\log\frac{\gamma^{c_{l_j}}}{\gamma^{c_{l_j}-}}\\
		&=\sum_{i=1}^ry_i\sum_{j=1}^s\left[L^{c_i}(l_j)-L^{c_i-}(l_j)\right]\log\frac{\gamma^{c_{l_j}}}{\gamma^{c_{l_j}-}}\\
		&=\sum_{i=1}^ry_i\sum_{l\notin T}L^{c_i}(l)\left\langle w,H^{c_l}\right\rangle\\
		&=\sum_{i=1}^ry_i\left\langle w,\sum_{l\notin T}L^{c_i}(l)H^{c_l}\right\rangle\\
		&=\sum_{i=1}^ry_i\left\langle w,L^{c_i}\right\rangle=\sum_{i=1}^ry_i\log\frac{\gamma^{c_i}}{\gamma^{c_i-}}.
	\end{align*}
	This completes the proof of \eqref{1} and thus completes the proof of the transient fluctuation theorem.
\end{appendices}

\setlength{\bibsep}{5pt}
\footnotesize\bibliographystyle{nature}

\end{document}